\documentclass{nature}
\usepackage{amsmath,amssymb, amstext}
\usepackage[T1]{fontenc}
\usepackage[latin1]{inputenc}
\usepackage{graphicx}
\usepackage{amssymb}
\usepackage{dcolumn}
\usepackage{color}
\usepackage{comment}
\usepackage{soul}
\usepackage{bm}
\usepackage{siunitx}
\usepackage[normalem]{ulem}
\usepackage{scalerel}
\usepackage{soul,xcolor}
\setstcolor{red}
\newcommand{\kk} {\ensuremath{{\bf k}}}

\newcommand{\QQ} {\ensuremath{{\bf Q}}}

\usepackage{url}
\usepackage[ colorlinks = true,
linkcolor = red,
urlcolor  = red,
citecolor = blue,
anchorcolor = red]{hyperref}
\usepackage[colorinlistoftodos]{todonotes}
\usepackage{bibentry}
\usepackage{pdfpages}

\begin{document}
	
	\title{Evidence of a compensated semimetal with electronic correlations at the CNP of %in magic angle 
	twisted double bilayer graphene}
	
\author{Ayan Ghosh$^{1}$\footnote{equally contributed}, Souvik Chakraborty$^{1}$\footnote{equally contributed}, Unmesh Ghorai$^{2}$\footnote{equally contributed}, Arup Kumar Paul$^{1}$, K. Watanabe$^{3}$,T. Taniguchi$^{3}$, Rajdeep Sensarma$^{2}$\footnote{sensarma@theory.tifr.res.in} and Anindya Das$^{1}$\footnote{anindya@iisc.ac.in}}
	
	\maketitle
	
	\begin{affiliations}
        \item Department of Physics, Indian Institute of Science, Bangalore, 560012, India.
          
        \item Department of Theoretical Physics, Tata Institute of Fundamental Research, Mumbai, 400005, India.
          
		\item National Institute for Materials Science, 1-1 Namiki, Tsukuba 305-0044, Japan.
	\end{affiliations}
	
	\noindent\textbf{Recently, magic-angle twisted bilayer graphene (MATBLG) has shown the emergence of various interaction-driven novel quantum phases at the %seq
	commensurate fillings of the moir\'e superlattice, while the charge neutrality point (CNP) remains mostly a vanilla insulator. Here, we show an emerging phase of nearly compensated semimetallicity at the CNP of twisted double bilayer graphene (TDBLG), a close cousin of MATBLG, with \textcolor{black}{signatures of electronic correlation}. 
	Using electrical and thermal transport, we find almost two orders of magnitude enhancement of the thermopower in magnetic fields much smaller than the extreme quantum limit, accompanied by a large magnetoresistance ($\sim 2500\%$) at CNP. This provides indisputable experimental evidence that TDBLG near CNP is a compensated semimetal. Moreover, at low temperatures, we observe an unusual %non-Fermi liquid 
 sublinear temperature dependence of resistance. A recent theory~\cite{Unmesh2022} predicts the formation of an excitonic metal near CNP, where small electron and hole pockets coexist. We understand the sublinear temperature dependence in terms of critical fluctuations in this theory.}
 
 %The electronic correlations determine the fate of the co-existing small electron and hole pockets near the CNP of the semimetallic TDBLG. Our theory with the formation of excitonic metal explains} \st{the resistance feature and} \textcolor{blue}{its unique sublinear temperature dependence.}
	
	%exciton formation %with finite momentum 
	%explain the feature, and the scattering of electrons and holes by critical Landau-damped excitonic fluctuations captures the unique sublinear temperature dependence.}

\begin{figure*}
		\centerline{\includegraphics[width=1\textwidth]{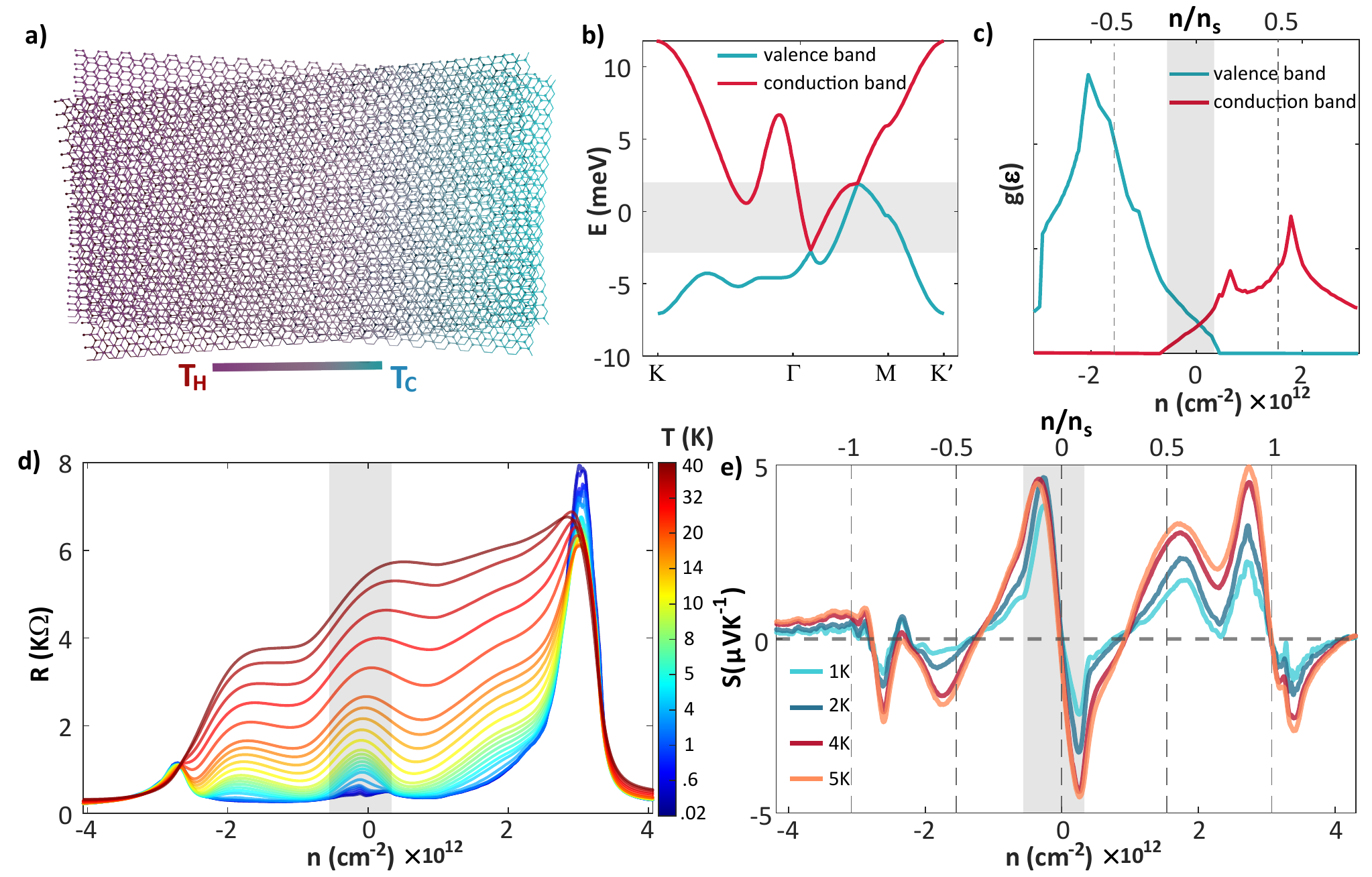}}
		\caption{\textbf{Thermal and electrical transport in TDBLG at zero magnetic field.} (a) Schematic of twisted double bilayer graphene with temperature gradient. (b) Band dispersion of TDBLG with twist angle $ 1.2^ o$ along the high symmetry axis in the mBZ. The band structure shows an energy overlap (shaded region) of $\sim4.5 \;\textrm{meV}$ between valence (blue) and conduction (red) bands. (c) Density of states of the valence (blue) and conduction (red) bands as a function of normalised carrier density ($n/n_s$). The shaded region indicates the presence of band overlap. (d) Resistance ($R$) versus carrier density ($n$) as function of increasing temperature. Several resistance peaks can be observed at full-filling of band and near the charge neutrality. The shaded region indicates the presence of band overlap. %Beyond $\sim 14K$ correlated peaks emerge at around $n/n_s \sim \pm 0.5$.
		(e) Evolution of thermopower as a function of normalized density $n/n_s$ at several temperatures. %\textbf{Evolution of thermopower with magnetic field :}
		}
		\label{Figure1}
	\end{figure*}

	\noindent\textbf{Introduction:}\
	The field of twist angle engineered moir\'e heterostructures have emerged as the latest platform to study strongly
	correlated quantum matter in condensed matter physics. Recent advances in controlled studies of graphene-based 
	moir\'e systems have unveiled a vivid spectrum of correlation-driven unconventional phases. For example, 
	in magic-angle twisted bilayer graphene (MATBLG), exotic phases and phenomena like superconductivity~\cite{cao2018unconventional,guinea2018electrostatic,Lu_efetov2019,yankowitz2019,cao2020nematicity,chichinadze2020nematic,liu2021tuning}, 
	correlated-insulator~\cite{Lu_efetov2019,liu2021tuning,cao2018correlated,tsai2019correlated,choi2019electronic,xie2019spectroscopic,saito2020independent,choi2021correlation}, 
	Chern-insulator~\cite{wu2021chern,das2020symmetry,stepanov2020competing,nuckolls2020strongly}, 
	ferromagnetism~\cite{sharpe2019emergent}, 
	%flavour polarization of Dirac cones~\cite{wong2020cascade,zondiner2020cascade},
	Dirac revival~\cite{wong2020cascade,zondiner2020cascade,choi2021correlation}, 
	and giant thermopower at low temperatures~\cite{paul2022interaction,kommini2021very} have been observed. 
	Twisted double bilayer graphene (TDBLG) is another prominent member of graphene-based moir\'e heterostructures, 
	where two sheets of Bernal-stacked bilayer graphene are stacked on top of each other with a controlled small twist angle between them (Fig.~\ref{Figure1}(a)). The resultant reconstruction of electronic levels into bands in the moir\'e Brillouin zone (mBZ) leads to the formation of low energy bands, 
	whose bandwidth is sensitive to the twist angle~\cite{Koshino2019,Chebrolu2019,Mohan2021}. The bandwidth is minimum around an %magic 
	angle $\sim 1.2^\circ$. Unlike MATBLG,
	the flat bands in TDBLG survive over a broader range of twist angles ($1.1^o-1.35^o$), thus providing a robust foundation to study strong correlation effects~\cite{burg2019correlated}.  %with a gap 
	The band structure of TDBLG can also be tuned by a perpendicular electric field~\cite{choi2019intrinsic,Liu_kim_2020,Mandar2020_tdblg,Cao_2020_tdblg}, which can drive the system from a metallic to an insulating state with interesting topological properties. %However, in absence of perpendicular electric fields, TDBLG has shown plain vanilla metallic behaviour near CNP%in previous studies
	%~\cite{choi2019intrinsic,Liu_kim_2020,Mandar2020_tdblg,Cao_2020_tdblg}, which can be understood in terms of non-interacting electrons.
	%an inheritance from its constituent bilayer graphene counterpart. 

    Theoretical predictions have shown that in TDBLG the low energy valance and conduction bands, though separated in momentum space, overlap in energy, allowing the coexistence of electron and hole pockets %that coexist close to 
	near the CNP~\cite{Koshino2019,Chebrolu2019,Mohan2021}. The coexistence of electron-hole pockets %This 
	has far-reaching and fascinating implications like colossal magnetoresistance~\cite{PhysRevB.103.155144}, large non-saturating thermopower with applied %perpendicular 
	magnetic field~\cite{skinner2018large} (these are also seen in Dirac and Weyl semimetals~\cite{ali2014large,inohara2017large,skinner2018large,Feng2021,he2021large}). A similarly enhanced magnetoresistance ($\sim 200\%$) has been reported for semimetallic bismuth and graphite~\cite{shan_wen_semimetal,article}. However, experimental demonstration of compensated semimetallic phase or the coexistence of electron and hole pockets in TDBLG has not been reported so far. Electronic interactions in such compensated semimetals can lead to the formation of excitonic insulator\cite{HALPERIN1968115,KHVESHCHENKO2004323} driven by Coulomb attraction between the electrons and holes, or to exciton condensation in metallic~\cite{Li_exc_SF_2017, anshul_exc_cond, Wang_exc_cond_2019} backgrounds near the CNP ~\cite{Unmesh2022}. Experimentally, the effects of strong electronic correlations in TDBLG have been seen when the samples are subjected to strong perpendicular electric fields~\cite{Mandar2020_tdblg,Liu_kim_2020,choi2019intrinsic,Cao_2020_tdblg} or magnetic fields~\cite{burg2019correlated} at the commensurate fillings of moir\'e superlattice. However, in absence of these perturbations, TDBLG has shown~\cite{Liu_kim_2020,Shen_2020} plain vanilla metallic behaviour without any report of strong electronic correlations at CNP. 
	
	In this work, we present a comprehensive study of temperature, carrier density and magnetic field-dependent resistance, and %thermal transport 
	thermopower of TDBLG with a twist angle $1.2^o$. At zero magnetic field, the thermopower is almost zero around the CNP due to compensation from opposite charge carriers (electrons and holes). With the application of a small magnetic field, the thermopower at low temperatures ($< 3K$) increases rapidly by up to $100-400$ times before saturating to $10-15 \mu V/K$ within $\sim 0.25T$. Similarly, the magnetoresistance (MR) at the CNP increases quite rapidly with the application of a small magnetic field and saturates before $1T$ with an enhancement of $2500\%$. The enhancement of thermopower by two orders of magnitude and large MR at the CNP of TDBLG is quite striking and is not seen for MATBLG in our study. A two-band particle-hole asymmetric model of TDBLG band structure with the co-existence of electron and hole pockets\cite{Koshino2019,Chebrolu2019,Mohan2021} is invoked to qualitatively and semi-quantitatively explain the results. The compensated semimetal allows a quantum transport regime with small Hall angles~\cite{Feng2021, shan_wen_semimetal}, where the electric current is dominated by Drude dissipative processes; however, the thermal current is dominated by the %non-dissipative 
	spiraling of the charge carriers in crossed electric and magnetic fields. These two different mechanisms lead to a strong enhancement of thermopower and magnetoresistance. We note that our data provide the first clinching evidence for compensated semimetallic phase %presence of simultaneous electron and hole pockets
	near the CNP in TDBLG.

  Further, the metallic behaviour (resistance decreasing with decreasing temperature) around the compensated semimetallic phase is quite striking. Near the CNP, We find an unusual sub-linear temperature dependence of the resistance ($R \sim T^{\alpha}$, $\alpha \sim 0.67-0.83$) below $10K$, whereas the temperature dependence %reverts to the 
   is super-linear ($R \sim T^{\alpha}$, $\alpha \sim 2-2.5$ ) as we dope away from the region of co-existing electron and hole pockets on either side of CNP. Note that the temperature dependence of the resistance becomes linear around the CNP above $10K$. Here, we invoke the recent theory of excitonic metal ~\cite{Unmesh2022} at the CNP of TDBLG to explain our unusual temperature dependence of the resistance.
   
   %We show that the Coulomb attraction between the small electron and hole pockets at the CNP forms an excitonic metal, which captures \st{the resistance feature and} its unusual non-Fermi liquid-like temperature dependence.  
   
   %The small Fermi pockets in this region are strongly affected by electronic correlations and the Coulomb attraction between electrons and holes lead to formation of excitonic condensates. The modified band-structure of this indirect excitonic metal can explain these resistance peaks qualitatively. 
   
   %Further, in this region near CNP, we find a unique sub-linear temperature dependence of the resistance ($R \sim T^{2/3}$) below $10K$. This non-Fermi liquid scaling of the resistance with temperature cannot be explained in terms of quasiparticles, which would normally produce super-linear temperature dependence; rather the scattering of the charge carriers in this critical metal by Landau-damped fluctuations of the excitonic order can explain this unique behaviour. The temperature dependence reverts to the usual Fermi liquid scaling ($R \sim T^2$) as we dope away from the region of co-existing electron and hole pockets on either side of CNP. Thus the presence of small Fermi pockets and associated strong correlations are crucial to the observation of this non-Fermi liquid phenomenon. We note that this is the first concrete evidence of presence of strong electronic correlations in TDBLG near CNP in absence of electric or magnetic fields.
	
\noindent\textbf{Device and setup:}\
The details of the device fabrication and measurement setup are mentioned in the supplementary information (SI-1). Here, we briefly mention it. The TDBLG device is realized by following the tear and stack technique~\cite{cao2018unconventional}. Two AB-stacked bilayer graphene sheets are stacked with a relative twist angle ($1.2^o$) and encapsulated by hexagonal boron nitride (hBN). For the two-probe resistance measurement, the standard lock-in technique ($13Hz$) with current bias has been used. For the thermopower measurement, an isolated gold line is placed parallel to our device, serving as a heater. Passing a current through the heater creates the temperature gradient across the device (Fig.~\ref{Figure1}(a)).  The generated thermometric voltage across the device is measured using standard $V_{2\omega}$ technique~\cite{zuev2009thermoelectric,PhysRevB.80.081413,nam2010thermoelectric,wang2011enhanced,duan2016high,ghahari2016enhanced,mahapatra2020misorientation}. For the temperature gradient, we have employed Johnson noise thermometry for the precise measurement of $\Delta T$, which gives an accurate value of Seebeck coefficient, $S = V_{2\omega}/\Delta T$. The Johnson noise thermometry is elaborated in our previous work~\cite{paul2022interaction} and shown in the supplementary information (see SI-2,4,5). All the measurements are performed in the linear regime (see SI-7). We note that for accurate measurement of the $\Delta T$, a linear fall-off of the temperature is necessary. Purposefully a simple two-probe geometry is implemented instead of having multiple metal leads (that can act as constant temperature heat sinks) which would have heavily altered the linear temperature falloff to a more complicated form. We have solved three-dimensional Fourier heat diffusion equations for a multi-layer stack using finite element calculations in Comsol (in SI-14) to test the validity of the assumption of a linear temperature profile in the case of a two-probe geometry. %For the same reason we are restrained from using dual gate geometry.  

\noindent\textbf{Electrical and Thermal Transport:}\ We first consider the carrier density-dependent %2-probe 
	resistance measured at several temperatures (between $20mK-40K$), shown in Fig.~\ref{Figure1}(d). The density, $n$, is obtained from the gate voltage applied to the $Si/SiO_2$ back gate by assuming an effective capacitance of the device. At low temperature, the resistance shows two strong peaks at large positive and negative densities of Fig.~\ref{Figure1}(d) with associated insulating behaviour (resistance increasing with decreasing temperature). We identify these peaks with the occurrence of moir\'e gaps, i.e. with densities $n =\pm n_s$, where the moir\'e conduction (valence) band of TDBLG is completely filled (emptied out). This corresponds to having $4$ electrons (holes) per moir\'e unit cell. Using the expression~\cite{cao2018unconventional,cao2018correlated} $n_s=\frac{4}{A}\approx \frac{8\theta^2}{\sqrt{3}a^2}$, where $A$ is the area of the moir\'e unit cell and $a$ is the graphene lattice constant; this translates to a twist angle of $\theta \simeq1.2^\circ$. This confirms that our sample has a twist angle within the range of observable flat bands for TDBLG~\cite{Liu_kim_2020,doi:10.1021/acs.nanolett.1c03066,Mandar2020_tdblg}.
        
    The resistance exhibits metallic behaviour at all densities other than the vicinity of the moir\'e insulator, including at the CNP. This is consistent with earlier work on magic angle TDBLG~\cite{burg2019correlated, Cao_2020_tdblg,Mandar2020_tdblg} and is in stark contrast with MATBLG, which behaves like an %correlated 
    insulator~\cite{cao2018correlated,yankowitz2019,Lu_efetov2019} near the CNP. Theoretically, this is explained by the fact that MATBLG shows protected Dirac nodes with zero density of electronic states at the Fermi level~\cite{BistritzerMacDonald2011}, while the valence and conduction bands of magic angle TDBLG overlap in energy~\cite{Koshino2019,Chebrolu2019,Mohan2021}, leading to formation of a compensated semimetal at the CNP with electron and hole pockets. The overlap of the valence (blue) and the conduction (red) bands of TDBLG can be seen from the theoretical band dispersions of TDBLG (Fig. \ref{Figure1}(b)), calculated at a twist angle of $1.2^\circ$. A detailed model of the moir\'e bands, which breaks particle-hole symmetry and includes trigonal warping, is needed to obtain this energy overlap of $\sim 4.5 ~\textrm{meV}$ between the bands (see SI-10 for details).  Fig.~\ref{Figure1}(c) shows the variation of the density of states (DOS) at the Fermi level with the carrier density $n$ (separately for these two bands). We see that the electron and hole pockets co-exist between carrier densities $n_-\sim -0.7 \times  10^{12}\; \textrm{cm}^{-2}$ ($n_-/n_s\sim-0.2$) and $n_+\sim  +0.4\times 10^{12}\; \textrm{cm}^{-2}$ ($n_+/n_s\sim0.12$). This density span is marked by grey shaded region in Fig.~\ref{Figure1}(c)-(e). The resistance at a fixed temperature shows additional peaks/dips as a function of density visible between the CNP and the moir\'e insulator. These features survive up to $\sim 20 K$ and become more prominent with increasing temperature. However, these features don't appear exactly at commensurate fillings and are not caused by strong electronic correlations; rather, they may be correlated with Van-Hove singularities crossing the Fermi level and are consistent with the reported results on TDBLG~\cite{burg2019correlated}. 

    We now focus our attention on thermal transport. The thermopower or Seebeck coefficient is defined as a generation of electric voltage due to a temperature difference ($S = -\frac{\Delta V}{\Delta T}$). Alternatively, using Onsager relation, $S$ can be written in terms of the Peltier coefficient (the ratio of heat current, $J_Q$ produced by an applied electrical current, $J_e$) as: $S = \frac{1}{T} \frac{J_{Q}}{J_{e}}$. At the CNP or in compensated semimetals, the heat carried by the opposite charge carriers %carrying the heat energy
	flows in opposite directions, and thus at a given applied current, $S$ will be proportional to: $\frac{J_{Q}}{J_{e}} \sim    \frac{(n_{e}-n_{h})}{(n_{e}+n_{h})}$ and will be very small. Here, $n_{e}$ and $n_{h}$ are the carrier concentrations for electron and hole, respectively. The density-dependence of thermopower at different temperatures from $1K$ to $5K$ is shown in Fig.~\ref{Figure1} (e). As expected, $S$ reverses sign at the CNP as well as at $n=\pm n_S$ (fully filled or completely empty moir\'e bands). %From the semiclassical definition, the sign of the thermopower depends on the charge of the majority carriers. For example, a hole (electron) like band gives positive (negative) thermopower, while it becomes zero at the symmetric points (like CNP) having both types of carriers. 
	We also observe several interesting features in the density dependence of thermopower away from CNP. These include additional change of sign at $n/n_s \sim +0.3, -0.4$, dips around $n/n_S \sim \pm 0.75$ and peaks in between. The sign change is related to change of topology of the Fermi surface (from electron (hole) like to a hole (electron) like) while peaks and dips may be related to possible Lifshitz transitions. However, in this letter, we focus on transport near CNP, leaving the explanation of these features for future work.

\begin{figure*}
		\centerline{\includegraphics[width=.9\textwidth]{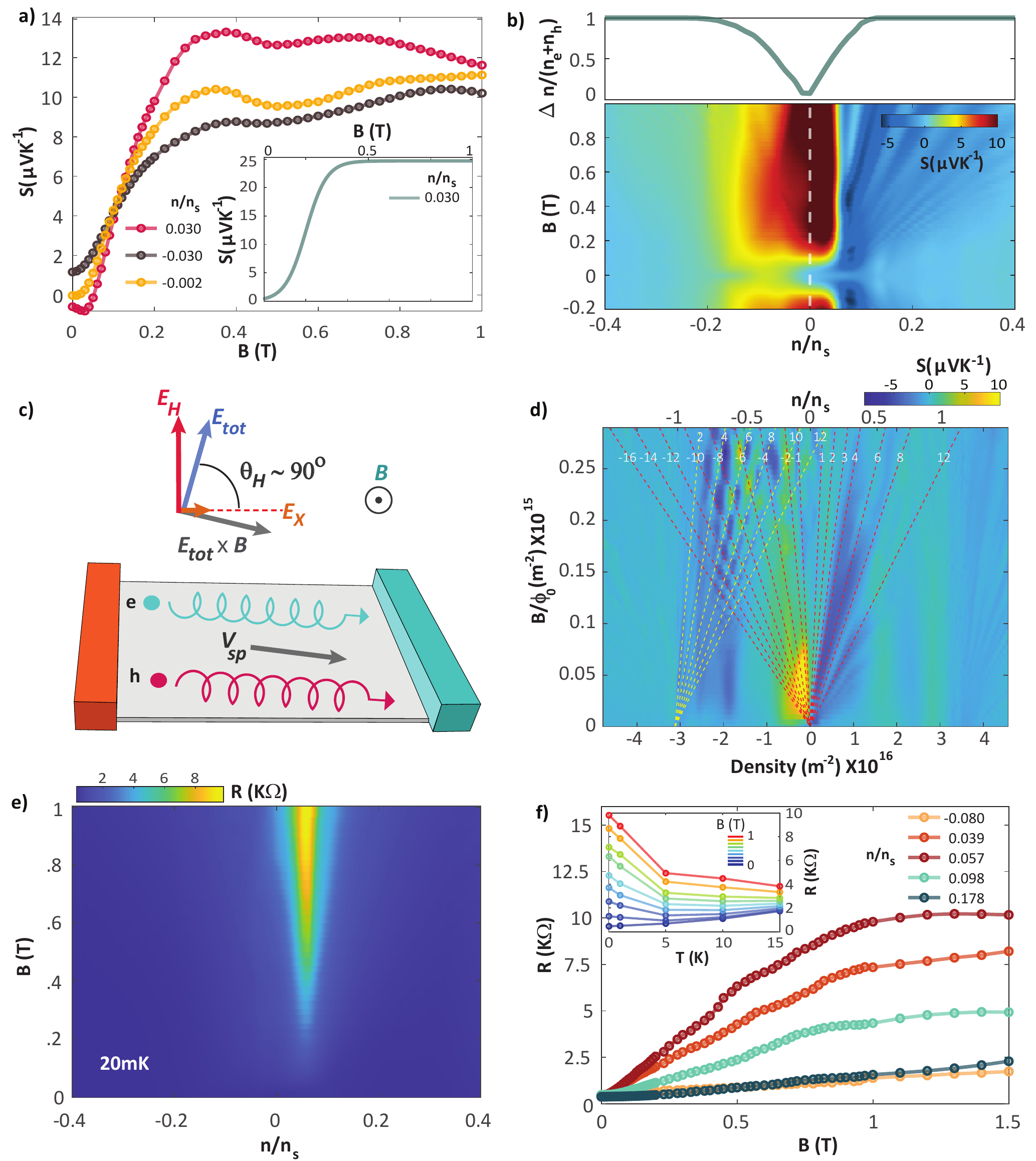}}
		\caption{ \textbf{Finite magnetic field thermopower and resistance measurements.} (a) Magnetic field dependence of measured thermopower at $n/n_s = + 0.03, -0.002, -0.03$ at $1K$. The inset shows the theoretically predicted thermopower for compensated semimetallic band. (b) (Top panel) Theoretically extracted normalized effective charge density ($\frac{(n_e-n_h )}{(n_e+n_h)}$) as a function of $n/n_s$. (Bottom panel) 2D color plot of thermopower as a function of perpendicular magnetic field and $n/n_s$ at $1K$. (c) Cartoon illustration of the (\textbf{E$_{tot}$} $\times$ \textbf{B}) drift %($\textbf{V}_\textbf{d}$) 
		on the carriers in the limit when $B$ approaches $B_H$ (or $\theta_H \rightarrow 90^o$). The Hall voltage along the y direction leads to an electric field  \textbf{E$_{H}$}, where $\lvert$\textbf{E$_{H}$}$\lvert \gg \lvert$\textbf{E$_{x}$}$\lvert$ in the limit $\theta_H \approx 90^o$.
		Electrons (labeled e) and holes (labeled h) both drift alongside in presence of crossed electric (\textbf{E$_{tot}$ = E$_{x}$}+\textbf{E$_{H}$ $\approx$ E$_{H}$}) and magnetic field (\textbf{B}) contributing additively to the heat current~\cite{Feng2021}. %[A more in depth illustration at different regimes is shown by Feng et. al~\cite{Feng2021}]
        (d) \textcolor{black}{Thermopower plotted over a  wide range of density/ $n/n_s$ (bottom axis / top axis) and magnetic field  (divided by flux quanta $\Phi_0$) with red (yellow) dashed lines marking the landau levels emanating from $n/n_s=0 (-4)$.}
        (e) \textcolor{black}{Resistance as a function of perpendicular magnetic field and $n/n_s$ at $20mK$.}
		(f) Perpendicular magnetic field dependence of measured resistance at several $ n/n_s $ at $ 1K $. The presence semimetallic band is further reflected through metal-insulator transition of resistance versus temperature curve with increasing perpendicular magnetic field, as demonstrated in the inset.
  }
		\label{Figure2}
	\end{figure*}

\noindent\textbf{Thermopower enhancement with low-magnetic field:}\ The metallic resistivity, together with the sign change of thermopower at CNP, strongly suggests that the system is a compensated semimetal. However, electrons and holes have opposite charges and respond to magnetic fields in different ways. Hence, to see clear evidence of ambipolar transport, we now consider thermal transport in our sample in the presence of a magnetic field $B$ applied perpendicular to the plane of the sample. Fig.~\ref{Figure2}(a) shows the variation of the measured $S$ with $B$ for three values of doping close to CNP at $1K$. From the figure, we find that $S$ is almost zero around the CNP in the absence of magnetic fields; it increases rapidly with the application of few milli-Tesla ($mT$) magnetic field and saturates around $\sim 10-14 \mu V/K$ beyond $0.3T$. The 2D-colour plot in the bottom panel of Fig.~\ref{Figure2} (b) shows that this large enhancement of thermopower is restricted to the vicinity of the CNP. 

    The enhancement of $S$ at relatively low-magnetic fields for compensated semimetals with electron and hole pockets can be understood in the following way, which has been quantitatively explained by Feng et. al~\cite{Feng2021}. In presence of crossed electric (E) and magnetic (B) fields, the drift velocity of charged particles have two components: a %dissipative 
    Drude response, $v_d=\pm \frac{\mu E }{1+(\mu B)^2}$ ($\mu$ being the mobility of electrons and holes) and a %non-dissipative 
    $\vec{E} \times \vec{B}$ spiralling component (shown in Fig.~\ref{Figure2}(c)). %which is independent of $B$. 
    For the electric current, the Drude response $\sim n_T=n_e+n_h$, while the response from the spiral component $\sim \Delta n= n_e-n_h$, where $n_e$ and $n_h$ are electron and hole densities. Since $\Delta n \ll n_T$, one can have a situation where the transport is in the quantum regime~\cite{Feng2021} ($\mu B \gg 1$) for $B >B_1$ (\textbf{$B_1=1/ \mu$}), but the Hall angle $\tan \theta_H = \frac{\sigma_{xy}}{\sigma_{xx}}= \mu B \Delta n/n_T$ is still small due to the compensation from the ratio of densities; i.e. the electric transport is dominated by the Drude response. However, for the thermal current, the Drude response $\sim \Delta n$, while the response from the spiral component $\sim n_T$, and hence thermal transport is dominated by the drift coming from the spiral component. In this regime, the thermopower $S_{xx}=J_Q/TJ_e$ is given by $S_{xx} \sim \frac {k_B^2T}{e\epsilon_F} \frac {\Delta n}{n_T } \mu^2 B^2$. 
%\begin{equation}
%	S_{xx} \sim \frac {k_B^2T}{e\epsilon_F} \frac {\Delta n}{n_T } \mu^2 B^2
%\end{equation}
    This rapid quadratic rise of the thermopower is seen in our data in Fig.~\ref{Figure2} (a). At higher magnetic fields $B>B_H$ (\textbf{$B_H\approx B_1\frac {n_T}{\Delta n}$}), where $\mu B \gg n_T/\Delta n$, one enters the regime of extreme quantum transport with large Hall angles, where the electric current is also dominated by the drift coming from the spiral component. In this case, the thermopower saturates and is given by %as a function of the magnetic field 
    $S_{xx} \sim \frac {k_B^2T}{e\epsilon_F} \frac {n_T}{\Delta n}$. The inset of Fig.~\ref{Figure2} (a) shows the theoretically
    calculated (mentioned in SI-11) thermopower for compensated semimetallic band~\cite{Feng2021}, which resembles very well with our experimental data.
%\begin{equation}
 % S_{xx} \sim \frac {k_B^2T}{e\epsilon_F} \frac {n_T}{\Delta n}
%\end{equation}
    In the top panel of Fig.~\ref{Figure2}(b), we plot $\left\vert\frac {\Delta n}{n_T }\right\vert$, obtained from theoretical band dispersions, as a function of the carrier density. We find that this theoretical $\left\vert\frac {\Delta n}{n_T }\right\vert$ falls to zero ($n_e$ $\approx$ $n_h$, hence $\Delta n= n_e-n_h\rightarrow0$) in the region where the large saturation value of thermopower is seen, corresponding %roughly 
    to the region where the electron and hole pockets are seen theoretically. Outside this region both the theoretical $\left\vert\frac {\Delta n}{n_T }\right\vert$ and the experimental data on thermopower are independent of carrier densities. These features provide further evidence that the thermopower enhancement near the CNP is due to the presence of simultaneous electron and hole pockets in the system. %We note that with magnetic field the sign reversal of $S_{xx}$ shifts from the zero density (Fig.~\ref{Figure2} (a) and Fig.~\ref{Figure2} (b)), and we attribute this to the different mobilities for electron and hole bands.

    Although the enhancement of thermopower for semimetals like bismuth~\cite{PhysRevB.14.4381,shan_wen_semimetal} and  tantalum phosphide~\cite{han2020quantized} in the bulk form have been reported earlier, the predicted saturation of thermopower with the magnetic field has not been observed experimentally so far. Our work on TDBLG provides the first experimental evidence of the saturation of thermopower for compensated semimetals accurately; we also demonstrate the tunability of the thermopower with carrier concentration because of the two-dimensional nature of our system. We would also like to mention that for comparison we have studied the thermopower response for MATBLG with the magnetic field, and it barely changes around the CNP (SI-8), as expected for non-compensated semimetals. It should be noted that the enhancement decreases as we increase the temperature and vanishes beyond $10K$ (SI-9) where quantum effects are destroyed due to increased scattering rate and the idea $\mu B \gg 1$ is no longer valid.

   We also have measured the thermopower over a wider range of filling and magnetic field (at $1K$) as shown in Fig.~\ref{Figure2}(d). Alongside the previously mentioned thermopower enhancement around CNP we also observe clear signatures of Landau fans emanating from $n/n_s=0,-1$. Using the famously known Diophantine equation \cite{dean2013hofstadter} for Landau levels (LL) we find a two-fold degenerate LL sequence at both fillings at high magnetic fields. We observe a further decrement in thermopower to occur beyond 2T. At these higher values of the applied perpendicular magnetic field, symmetry breaking causes a gap opening at CNP. The lack of DOS (due to this gap opening) causes decay in thermopower with any further increase in a magnetic field. At even higher fields, Landau levels start emerging resulting in oscillations in thermopower  along the crossings of the Landau fans.

\noindent\textbf{Large magnetoresistance at CNP:}\ 
	To further investigate the unambiguous footprints of electron and hole pockets in TDBLG, we examine the magnetoresistance (MR) of the system and its temperature dependence near the compensated region. In  Fig.~\ref{Figure2}(e) a large enhancement of MR (measured at $1 K$) confined within the vicinity of \textcolor{black}{CNP} can be observed in the 2D-colour plot of R as a function of $B$ and $n/n_s$. %Figure~\ref{Figure2}(d) focuses on the finite magnetic field resistance measurements. %The field direction in these measurements is always perpendicular to the sample plane. 
    Fig.~\ref{Figure2}(f) shows resistance (measured at $20 mK$) as a function of the magnetic field at different carrier densities. Close to CNP, the MR monotonically increases with $B$ and saturates around a magnetic field of $1T$ with a maximum increment of $2500\%$.  The rapid rise of the MR with a magnetic field can be understood from the quantum limit of electrical transport~\cite{kumar2017extremely,skinner2018large}, where it is still dominated by the Drude response, while the saturation behaviour is dominated by the %non-dissipative
    spiral component of current. We note that the high MR and the saturation is a  distinctive behaviour of many compensated semimetals~\cite{shan_wen_semimetal,liang2015ultrahigh}. Normal metals, on the other hand, have higher scattering rates which limit their magnetoresistance. 
	The temperature dependence of resistance at various magnetic fields at a fixed density near the charge neutrality is shown in the inset of Fig.~\ref{Figure2}(f). The behaviour is very similar to that of previously reported compensated semimetals~\cite{PhysRevB.65.241101,shan_wen_semimetal,article}. We can clearly see that the system exhibits a field-induced metal-to-insulator transition around a magnetic field of $0.2T$.
	
 %The charge density wave order degrades as one moves away from the CNP on either side, as the mismatch between the size of the electron and hole pockets grows, and naturally vanishes when the system has only one type of carrier.
    
    \begin{figure*}
		\centerline{\includegraphics[width=1\textwidth]{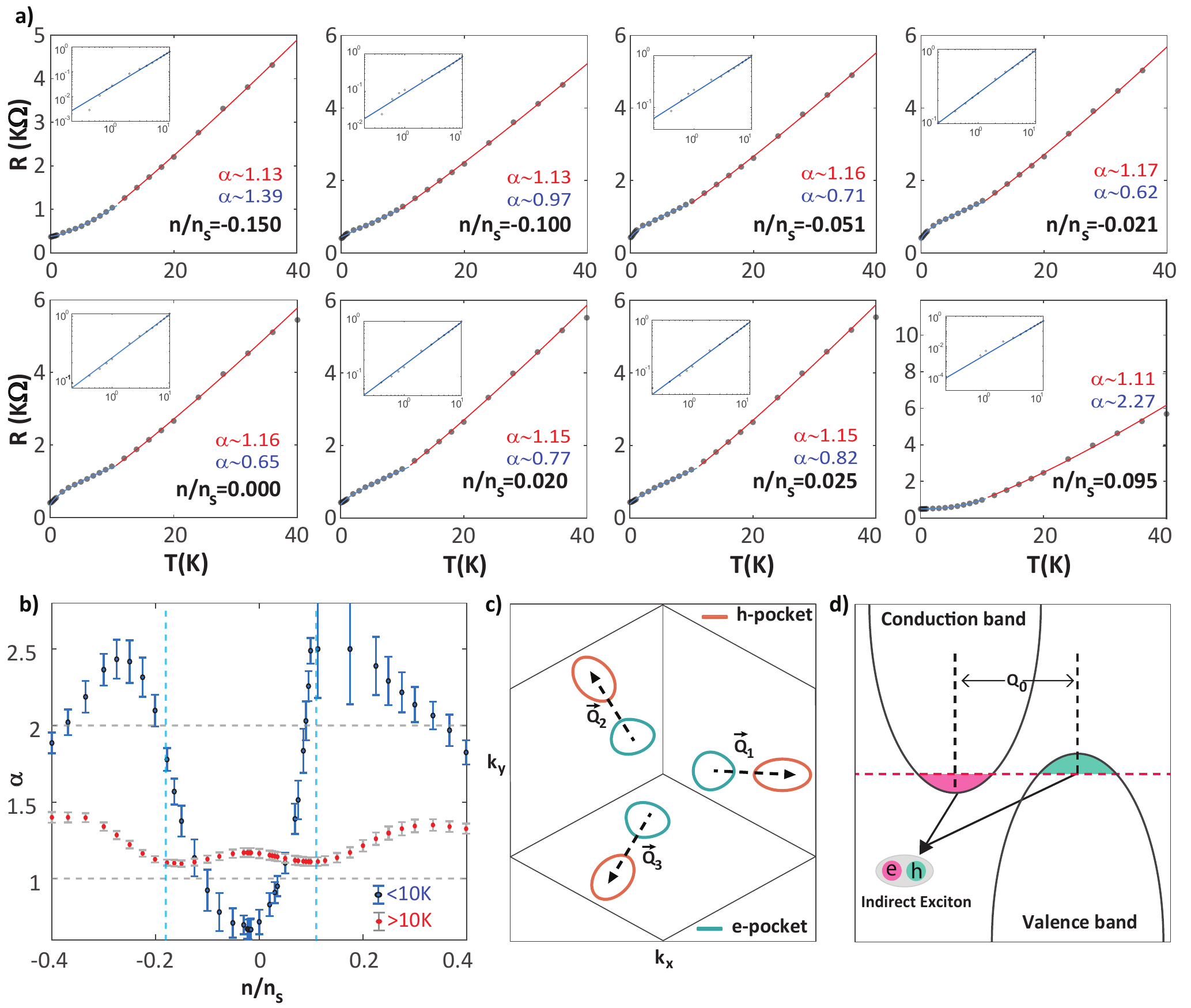}}
		\caption{\textbf{Electronic correlations in TDBLG near CNP:.}
  \textcolor{black}{
        (a) Power-law fitting of resistance as a function of temperature at different densities. Away from the CNP, at  $n/n_s \sim -0.150,-0.100,0.095$, the best fit for $R = R_0 + AT^\alpha$ is shown as the blue (red) line at low temperature (high temperature) range. The low temperature exponent shows a clear super-linear behaviour. In contrast near CNP, at $n/n_s \sim -0.05, -0.02, 0.0, 0.02$, the resistance shows a sub-linear behaviour (blue line is the fit) at low temperature ($0.02-10~K$). The insets show the data in log-log scale. The behaviour of $R$ above $10 ~K$ at all density ranges is almost linear with $T$.
        (b) Fitted exponent ($\alpha$) both for low temperature ($<10K$) and higher temperature fitting ($>10K$ and $<40K$)  depicted by black circles (with blue error bars) and red  circles (with grey error bars) respectively, the blue vertical dashed lines frames the region of electron and hole coexistence. The low temperature exponent shows nontrivial sub-linearity inside the aforementioned coexistence.} (c) Calculated Fermi surfaces in TDBLG at CNP showing the electron (blue) and the hole (red) pockets. Note that the centres of the pockets are shifted in momentum space. (d) Schematic representation of indirect excitons as pairing between electrons and holes. 
		}
		\label{Figure3}
              \end{figure*}
              
\noindent\textbf{Sub-linear temperature scaling of resistance around CNP:}\  As seen in Fig.~\ref{Figure1}(d), we observe a monotonic increment of resistance with temperature in the whole
        density range, suggesting metallic transport throughout the flat band. In Fig.~\ref{Figure3}(a), we study the temperature dependence
        of the resistance of the sample at several densities on either side of CNP. At large positive or negative densities, when there is only one type of carrier, we see that the resistance has a super-linear (with exponent between $2-2.5$) behaviour with temperature below $10K$ (marked by blue solid line). However, the situation changes dramatically near CNP, where both carriers are present. The resistance at these densities ($n/n_s=-0.051,~-0.021,~0.00,~0.02$) show a sub-linear behaviour with temperature in the range $200mK-10K$ marked by blue solid line, and a linear dependence above $10K$. The sub-linear dependence can be fitted using ($R=aT^\alpha+R_0$), and obtain $\alpha$ between $0.67-0.83$. The insets shows the low temperature \textcolor{black}{regime} (blue fitting) in a log-log scale to better represent the low temperature behaviour. The evolution of $\alpha$ with $n/n_s$ is shown in Fig.~\ref{Figure3}(b) (in blue circles with error-bar). It can be noticed from Fig. 3b that the sub-linear temperature dependence is prominent in the region where the electron and hole pockets co-exist, which is marked by the vertical dashed lines in Fig. 3b.  %$\alpha \sim 2/3$ in this regime (fitted $\alpha$ varies between $0.67-0.83$). 
        Note that as described before, in this measurement scheme (optimized for thermopower) we are limited to two-terminal resistance measurements, which has finite contributions from the contact resistance. It should be noted that beyond the full-filling in Fig. 1d, the resistance barely changes with increasing temperature,  where as within the full-filling the resistance increase monotonically (metal-like) with increasing temperature (Fig. 3a and Fig. 1d) and changes by $\sim 200 \Omega$/Kelvin. This suggests that the contact resistance at the measured low temperatures range ($200mK-10K$) barely changes, and our data is predominantly governed by the channel resistance. 
        %This is supported by the monotonic (metal-like) increase  of resistance with the temperature at large densities ($n\approx \pm 2.58 \times 10^{12} cm^{-2}$) which is contrary to the more common insulator-like temperature dependence in case of large contact resistance values with strong dependence on temperature.
        %We note that the sublinear scaling of $R$ with $T$ extends in temperature above the $T_c$ of the excitonic order and cannot be explained within a Fermi liquid picture of almost non-interacting quasiparticles.

        The observed sub-linear temperature dependence of the resistance around the CNP in Fig. 3a and 3b is quite unusual. Note that in graphene~\cite{RevModPhys.82.2673} and MATBLG~\cite{Lu_efetov2019,paul2020interplay}, one finds insulating behaviour near the CNP due to vanishing density of states. In contrast, small electron and hole pockets are formed in TDBLG near CNP, (see Fig.~\ref{Figure3}(c)) which leads to metallic behaviour. However one would expect a $T^2$ behaviour of resistance at the lowest temperatures and a linear temperature dependence above Bloch Gruneissen temperature, which is within a few Kelvin in TDBLG near CNP. Indeed, we find linear dependence of resistivity for $T>10 K$. Though, beyond $n/n_s=\pm 0.2$ the fitting tends to move marginally away from the expected linearity as seen in Fig. 3b. It should be noted that, beyond $n/n_s=\pm 0.2$ the Fermi surface increases resulting in higher Bloch Gruneissen temperature, as a result to see the expected linearity one may need to go higher temperature than the measured temperature range. %At high desnities it is important to note that due to the super linearity it is very difficult to distinguish the superlinear scaling and the linear scaling transition point. The full temperature range ($200mK-40K$) can be interpreted as a singular superlinear fitting with power closely matching $\alpha\sim2$. Additionally, beyond $n/n_s=\pm 0.2$ the fermi surface increases resulting in higher Bloch Gruneissen temperature as a result the expected linear R-T behaviour may not match the fitted temperature range.} 
        To understand the sub-linear dependence around the CNP, we rely on the recent theory ~\cite{Unmesh2022}, which predicts the formation of excitonic condensate due to Coulomb attraction between the electron and hole pockets. Since the electron and hole pockets are shifted in momentum (see Fig.~\ref{Figure3}(c)), indirect excitons with momenta connecting the center of the pockets are formed, as shown in Fig.~\ref{Figure3}(d). This leads to an excitonic metal at low temperatures. The fermions in this metal are scattered by Landau damped critical fluctations of the excitonic order. This leads to a non-Fermi liquid behaviour~\cite{Unmesh2022,SachdevMetlitski,Metlitski2,Polchinski,Chubukov,sslee}, where the scattering rate and hence the resistance, $R\sim T^{2/3}$. Note that the hole pocket, when shifted by the momentum of the exciton, lies on top of the electron pocket and hence this is similar to the scenario for an order parameter with zero momentum~\cite{SachdevMetlitski}, which is known to lead to $T^{2/3}$ scaling of scattering rates~\cite{Unmesh2022}. Based on the theory, we believe that our experimental data with sub-linear temperature dependence of the resistance with exponent $0.67 - 0.83$ around the CNP shows the signature of excitonic metal in TDBLG.
         %We note that this is the first time such a strongly correlated non-Fermi liquid behaviour is being reported in TDBLG. 
	
%\noindent\textbf{Discussion:}\
	    In summary, we have reported strong enhancement of thermopower and magnetoresistance in TDBLG at low temperatures near the charge neutrality point for relatively modest magnetic fields. This behaviour is understood in terms of electric and heat transport in a compensated semimetal and provides unambiguous evidence of the presence of simultaneous electron and hole pockets in this system. The resistance at low temperatures shows  a sublinear dependence, which has been attributed to the formation of an excitonic metal described in recent theoretical work~\cite{Unmesh2022}. \textcolor{black}{It will be interesting to see how these features evolve with a perpendicular displacement field, which is left for studies in the future.} 
     
     %in this system due to Coulomb interaction between the electron and hole pockets. %The strong electronic correlations drive the system into a non-Fermi liquid state, which shows sublinear increase of resistance with temperature. %, $R\sim T^{2/3}$. We thus present the first evidence of strong electronic correlations in a TDBLG system near the charge neutrality point at zero perpendicular electric or magnetic field.
	    \newpage
\noindent\textbf{Methods:}\

\subsection{Device fabrication and measurement setup:}
For assembling the hBN encapsulated TDBLG, we have used the standard `tear and stack' technique~\cite{cao2018correlated,cao2018unconventional}. The encapsulated device is placed on a $Si/SiO_2$ substrate acting as an electrostatic gate. The fabrication process is explained in much grater details in SI (SI-1). The length and width of the representative device are approximately 6 $\mu m$ and 3 $\mu m$, respectively. An optical image of the measured device is provided in SI-fig. 3. An isolated thin gold line, placed $\sim 3 \mu m$ away from the source probe acts as a heater. During thermopower measurement, upon injecting a current ($I_\omega$) in the heater line a temperature gradient arise  across the length of the device. The source contact neighbouring the heater gets hotter ($T_h$) while the drain is maintained at constant bath temperature of the mixing chamber (m.c) plate due to cold ground. The voltage ($V_{2\omega}$) generated across the channel is measured using standard Lock-in amplifier (SI-2). For resistance measurement low-frequency ($\sim 13 Hz$) Lock-in technique (SI-2) is employed. To measure the temperature difference ($\Delta T$), we employ Johnson noise thermometry. The noise thermometry circuit consists of LC resonant ($f_r\sim 720kHz$) tank circuit, followed by a cryogenic amplifier and a room temperature amplifier (see SI-fig. 2d). A detailed gain calculation of the amplifier chain is mentioned in the SI (SI-6). As depicted in SI-fig. 2a a relay situated on the mixing chamber plate  is used to switch between high-frequency ($\Delta T$) and low-frequency (Resistance and $V_{2\omega}$) measurement scheme.

\subsection{Theory:} 
Twisted Double Bilayer Graphene consists of two Bernal stacked (AB) bilayer graphene (BLG) sheets with a relative twist angle $\theta$ between them. Here, we work with the ABAB stacking, so that the $B$ sublattice of the top interface layer sits on top of the $A$ sublattice of the bottom interface layer. Here we consider the band structure of TDBLG following Ref.s~\cite{Koshino2019,Mohan2021}. The details of the Hamiltonian construction can be found in the SI (SI-10). For this work, we have taken the following coupling parameters ~\cite{Koshino2019},
	$\hbar v_0/a=2.1354~\text{eV}$ (the nearest neighbour tunneling amplitude along the monolayer graphene sheet), $ \gamma_1=400~\text{meV}$ (the c-axis inter-layer hopping between the dimer sites), $\gamma_3=320~\text{meV}$ (the inter-layer hopping between the non-dimer sites), $\gamma_4=44~\text{meV}$ (the coupling between dimer and non-dimer sites), and $\Delta'=50~\text{meV}$ (the potential difference between dimerized and non-dimerized sites). For the AA/BB and AB tunneling amplitudes across the twisted layers, we have used~\cite{Koshino2019}, $u=79.7~\text{meV}$ and $u'=97.5~\text{meV}$ respectively in our calculations. In this work, we have taken a 184 dimensional matrix which gives an error of $< 1\%$ in the band dispersions at the magic angle of $1.2^\circ$.

	The Coulomb attraction between the electron and hole pockets lead to formation of indirect exciton condensates in TDBLG near CNP. In this calculation we will replace the Coulomb potential between electrons and holes by a screened short range potential. In fact we will use an effective momentum independent potential with the energy scale $V_0\sim10.8\;\textrm{meV}$. Note that there are three electron pockets separated from the three hole pockets by wavevectors $Q_{1(2)(3)}$.
The mean field Hamiltonian describing the excitonic condensate is given by 
\begin{equation}\label{meanfham}
		\mathcal{H}(\QQ_i)=\begin{bmatrix}
		\frac{1}{3}\epsilon^c_{\kk}-\mu&\Delta\\ \Delta& \frac{1}{3}\epsilon^v_{\kk+\QQ_i}-\mu
		\end{bmatrix}
		\end{equation}
where, the $\epsilon^{c(v)}_{\kk}$ represents the non-interacting conduction (valence) band dispersion and the chemical potential is denoted as $\mu$. Note the order parameter $\Delta$ is same for all the pockets and is determined self-consistently. We can then write the modified quasi-particle dispersion relation in presence of the excitonic condensate,
      \begin{equation}
	E^{\pm}_{\QQ_i}(k)=\frac{\epsilon_\kk^c+\epsilon_{\kk+\QQ_i}^v}{6}-\mu \pm \sqrt{\frac{(\epsilon_\kk^c-\epsilon_{\kk+\QQ_i}^v)^2}{36}+\Delta^2}
	\label{disp_exc}
	\end{equation}
    The above energy spectrum generates a finite Fermi surface near CNP, which leads to metallic transport in presence of the condensate. We can use the inverse density of states calculated from this modified dispersion as a proxy for the resistivity of the material and this inverse DOS is plotted as a function of carrier density in the main text [Fig. 3(c) inset] to indicate the rough behaviour of the resistance in the system near CNP.

\noindent\textbf{References:}\
	\bibliography{ref}

\noindent\textbf{Acknowledgments:} A.D. thanks the Department of Science and Technology (DST) and Science and Engineering Research Board (SERB), India for financial support (DSTO-2051) and acknowledges the Swarnajayanti Fellowship of the DST/SJF/PSA-03/2018-19. K.W. and T.T. acknowledge support from the Elemental Strategy Initiative conducted by the MEXT, Japan and the CREST (JPMJCR15F3), JST. UG and RS acknowledge computational facilities at the Department of Theoretical Physics, TIFR Mumbai.  R.S. acknowledges support of the Department of Atomic Energy, Government of India, under Project Identification No. RTI 4002. 

\noindent\textbf{Author contributions:} A.G, S.C and A.K.P contributed to device fabrication and data acquisition. A.G contributed to analysis. A.D. contributed in designing the experiment, data interpretation and analysis. K.W and T.T synthesized the hBN single crystals. U.G, and R.S contributed in development of theory, data interpretation, and all the authors contributed in writing the. manuscript.

\noindent\textbf{Competing Interests:} The authors declare that they have no competing interests. 

\noindent\textbf{Data and materials availability:} All data needed to evaluate the conclusions in the paper are present in the paper and/or the Supplementary Materials. Additional data related to this paper will be available upon reasonable request to the author.



\section*{Supplementary material}

# SI: Evidence of a compensated semimetal with electronic correlations at the CNP of twisted double bilayer graphene

Ayan Ghosh[1][*], Souvik Chakraborty[1][†], Unmesh Ghorai[2][‡], Arup Kumar Paul[1], K. Watanabe[3], T. Taniguchi[3], Rajdeep Sensarma[2][§] and Anindya Das[1][¶]

[1]Department of Physics, Indian Institute of Science, Bangalore, 560012, India.
[2]Department of Theoretical Physics, Tata Institute of Fundamental Research, Mumbai, 400005, India
[3]National Institute for Materials Science, Namiki 1-1, Ibaraki 305-0044, Japan.

**SI-1: Sample preparation**

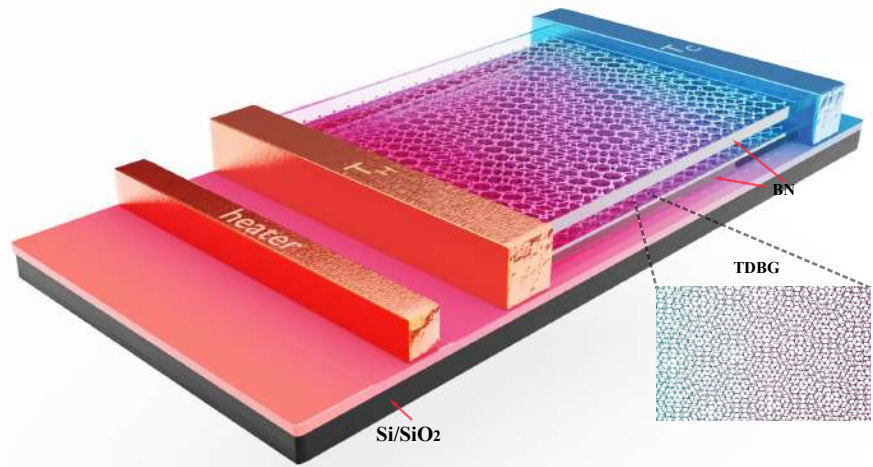

**Supplementary Figure 1:** (a) Cartoon depiction of a twisted double bilayer graphene encapsulated in $hBN$ on top of a Si/SiO$_2$ substrate along with source drain and heater contacts. The temperature gradient is shown through a colour gradient (hot junction is red while cold junction is blue)

Well established "tear and stack" [1]technique was employed to assemble the representative twisted double bilayer graphene (TDBLG) devices. As shown in SI-fig. 1, the hetero-structure consists of twisted double bi-layer graphene encapsulated in-between two $hBN$ layers. At first, $hBN$ ($\sim 10nm - 25nm$) and graphene are exfoliated on separate Si/SiO$_2$ substrates and suitable flakes are chosen under an optical microscope. Then, using a transparent PDMS-poly propyl carbonate (PPC) stamp [1] the top $hBN$ flake is picked up. This PDMS-PPC-$hBN$ stamp is then used to sequentially pick up graphene layers (torn from a single graphene sheet). The small twist angle is created by adding a small rotation to the rotation stage

[*]equally contributed

[†]equally contributed

[‡]equally contributed

[§]sensarma@theory.tifr.res.in

[¶]anindya@iisc.ac.in

(holding the substrate containing the unpicked graphene flake) in between two successive pickups of the individual graphene flakes. This is followed by another $hBN$ flake to encapsulate the TDBLG layers. The completed stack is then transferred on a clean Si/SiO$_2$ substrate kept at 75$^o$ C. During the graphene pick-up process, steady temperature between 50$^o$-55$^o$C is maintained to minimize the effect of thermal relaxation of the twist angle.

The samples are then spin-coated with 495A4-950A4 PMMA bilayer. After depositing each PMMA layer, sample is baked at 80$^o$C for 15 minutes, a rather departure from the usual 180$^o$C and 5 minutes bake time. This low-temperature baking ensures least thermal relaxation of the twist angle. Up next, the contacts, and the heater lines, are defined using standard e-beam lithography. After developing the treated photoresist, the contacts are etched with CHF$_3$-O$_2$ plasma. This is followed by thermal deposition of $Cr(2nm)$, $Pd(10nm)$ and $Au(70nm)$ and another etching to define the geometry of the TDBLG channel. The width of source and drain contacts on TDBLG are $1\mu m$, and the heater lines are $\sim 200-300 nm$. The heater lines are separated by a $3\mu m$ gap from the source contact.

## SI-2: Experimental measurement setup

**Supplementary Figure 2: Setup for the measurement:** (a) Schematic of the measurement setup. (b), (c) and (d) shows simplified form of the resistance, V$_{2\omega}$ and $\Delta T$ measurement schemes, respectively.

SI-fig. 2(a) displays the complete schematics for resistance, thermopower and thermal noise measurement setup. The section of the setup enclosed within the turquoise dashed box, holds the device on a chip-carrier located on the cold finger connected to the mixing chamber plate (MC plate). The device

contact far away from the heater is directly anchored to the cold finger and serves as a cold ground (c.g. coloured in turquoise). While the contact adjacent to the heater is used for the resistance, V$_{2\omega}$ as well as for $\Delta T$ measurement. The later contact, from here onwards will be addressed as hot contact. The hot contact branches out into two connections: one going to the left most coaxial (coax.) cable through a 1M$\Omega$ resistance, and another going to the relay, which serves to switch between the low frequency (resistance and V$_{2\omega}$) measurement scheme and high frequency (thermal noise) measurement scheme.

SI-fig. 2(b) shows a simplified schematics for resistance measurement. The left most coaxial line is used to inject low frequency (13 Hz) current (I$_{ac}$) through the ballast resistance (1$M\Omega$) into the sample. The voltage drop ($V_S$) across the sample is measured through the right most coaxial line using a SR560 voltage pre-amplifier followed by a lock-in amplifier. Here R$_s$ stands for the sample resistance.

SI-fig. 2(c) shows the simplified V$_{2\omega}$ measurement scheme. The two coaxial cables at centre, are used to inject ac (I$_h$) through the heater line 13 Hz via a $1k\Omega$ resistance connected in series to the heater line acting as the ballast resistance. Similar to the resistance measurement the Seebeck voltage or the second harmonic V$_{2\omega}$ signal generated across the sample is measured via the right most coaxial cable.

SI-fig. 2(d) shows the simplified schematic of the $\Delta T$ measurement setup which utilizes Johnson noise thermometry. The noise measurement setup consists of an LC resonant circuit ($f_r \sim 720 kHz$), followed by a low noise HEMT cryo-amplifier (cold amplifier), placed on the 4K plate (shown as the purple dashed box) inside the dilution refrigerator. A second stage amplification is further conducted by a room temperature amplifier (RTA) and then sent to a spectrum analyzer. During the excess thermal noise measurement the hot contact of the sample is switched to the input of the LC resonant circuit using the relay. The LC circuit acts as a band-pass filter and, at resonance, has unit gain. For noise measurement, equivalent dc is injected through the heater line to create the temperature gradient across the sample.

**SI-3: Estimation of twist angle**

**Twisted Double bilayer sample:** The twist-angle $\theta$ of the TDBLG device is determined from the densities at the full-filling $\nu = \pm 4$ (i.e 4 electron per unit cell of the moire superlattice) of the low-energy band. The area of superlattice unit cell for a given twist angle is given by $A \sim \sqrt{3}a^2/2\theta^2$, where $a = 0.246nm$ is the lattice constant of monolayer graphene. At the full-filling, the superlattice density $n_s$ is given by $n_s = 4/A = 8\theta^2/\sqrt{3}a^2$. Using this expression the $\theta$ is determined form the densities at $\nu = \pm 4$. For this device, the secondary Dirac peak at $\nu = +4$ is observed to occur at a density of $n = 3.05 \times 10^{12} cm^{-2}$ (SI-fig. 3), which translate into a twist angle of $1.2^0 \pm 0.01^0$.

**Magic angle twisted bilayer sample (MATBLG):** The twist-angle $\theta$ is calculated following procedure mentioned in our previous work [2]. For this device $\nu = \pm 4$ occur at densities $n = \pm 2.58 \times 10^{12} cm^{-2}$, translating to a twist angle of $1.05^0 \pm 0.2^0$.

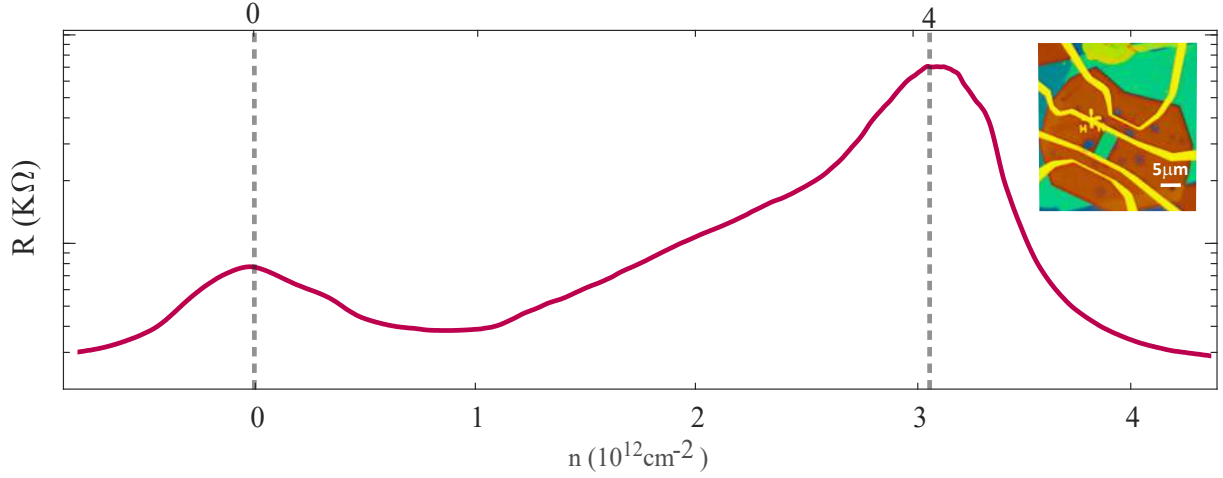

**Supplementary Figure 3:** Resistance versus density (n) plot for Twisted Double bi-layer graphene device at $2K$, showing multiple resistance peaks, the top axis is mapped to filling of the super-lattice unit cell. The inset shows the corresponding device image

### SI-4: Excess thermal noise ($S_V$ = 2k$_B\Delta$TR) vs. Temperature

We have employed Johnson noise thermometry to measure the temperature difference $\Delta T$ across the sample. [3, 4, 5, 6, 7, 8]. The sample temperature remains in equilibrium with the MC plate via cold ground in the absence of any heater current. In this case, The measured thermal noise by the cryo-amplifier is given by $S_V{}^0 = 4k_B T_c R_s$, where $T_c$ refers to the temperature of the MC plate and $R_s$ is the sample resistance. In presence of finite heater current, in the linear regime, the temperature profile across the TDBLG as a function of distance ($x$) from the hot contact is qualitatively depicted in SI-fig. 4. Here the position $x = 0$, corresponds to the junction of TDBLG channel and hot contact at temperature $T_h = T_c + \Delta T$, with $\Delta T$ being the net temperature difference between the hot and cold contacts. The position $x = L$, corresponds to the junction of graphene and cold contact at temperature $T_c$. Thus, at any position $x$ on the TDBLG the temperature is given by $T(x) = T_c + \Delta T(x)$, where $\Delta T(x) = \Delta T(1 - x/L)$. In this scenario, The total thermal noise seen by the amplifier is the cumulative sum of noise coming from all the infinitesimal sections along the length of the TDBLG channel. Hence, the measured noise ($S_V{}^I$) is given by:

$$S_V{}^I = 4k_B \int_0^L (T_c + \Delta T(x))\rho(x) dx \qquad (1)$$

where, $\rho(x)$ corresponds to the resistivity of the TDBLG at a distance $x$ into the channel. Considering homogeneous resistance i.e $\rho(x) = R_s/L$, the eqn. 1 can be rewritten as

$$S_V{}^I = 4k_B T_c R_s + 4k_B \rho \int_0^L (\Delta T(x)) dx$$
$$= 4k_B T_c R_s + 4k_B \rho \Delta T \int_0^L (1 - x/L) dx$$
$$= 4k_B T_c R_s + 2k_B \Delta T R_s$$

4

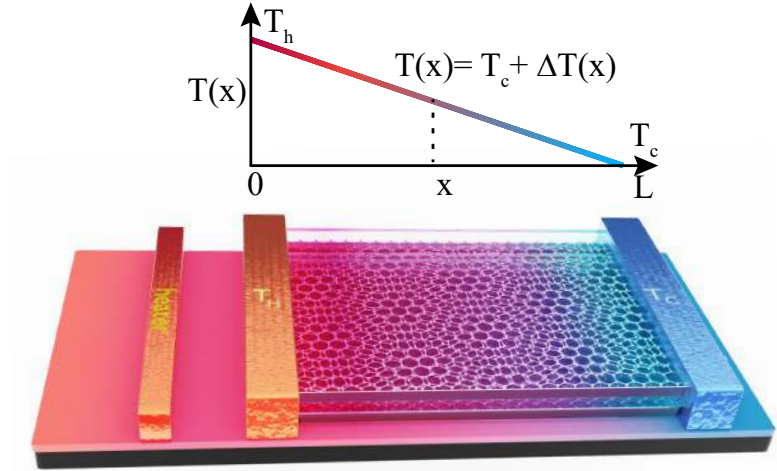

**Supplementary Figure 4:** Temperature gradient ($\Delta T$) across the device in presence of heater current ($I_h$).

Thus, the excess thermal noise due to applied temperature gradient is given by $S_V = S_V^I - S_V^0 = 2k_B \Delta T R_s$.

### SI-5: Measuring $\Delta T$ from excess thermal noise

The net $\Delta T$ is measured, by first measuring the thermal noise with varying the heater current from zero. The excess thermal noise is determined by subtracting the zero current thermal noise from the total measured noise at any given current. SI-fig. 5 shows the measured excess thermal noise ($S_V$) as function of square of applied heater current ($I^2$) at various temperatures. The linearity of the measured $S_V$ with $I^2$, in this plots, again corroborates the linear response regime. The temperature gradient $\Delta T$ is derived from the $S_V$ data using $S_V = 2k_B R \Delta T$ as shown in SI-fig. 6. The red lines are the gaussian smoothening of the scattered data showing linearity of $\Delta T$ with $I^2$.

SI-fig. 7(a) shows $\Delta T$ versus $I^2$ at $\nu = 0.4$ at $T = 1K$ with (red) and without (orange) perpendicular magnetic field. It can be seen that the measured $\Delta T$ does not depend on magnetic field. These observations indicate that $\Delta T$ is determined solely by the substrate. SI-fig. 7(b) shows $\Delta T$ versus $I^2$ inside the flat band $\nu = -3.5$ (orange) and near the full-filling $\nu = 4$ (blue) at $T = 1.2K$. As can be seen the measured $\Delta T$ does not depend on the carrier concentration of the sample.

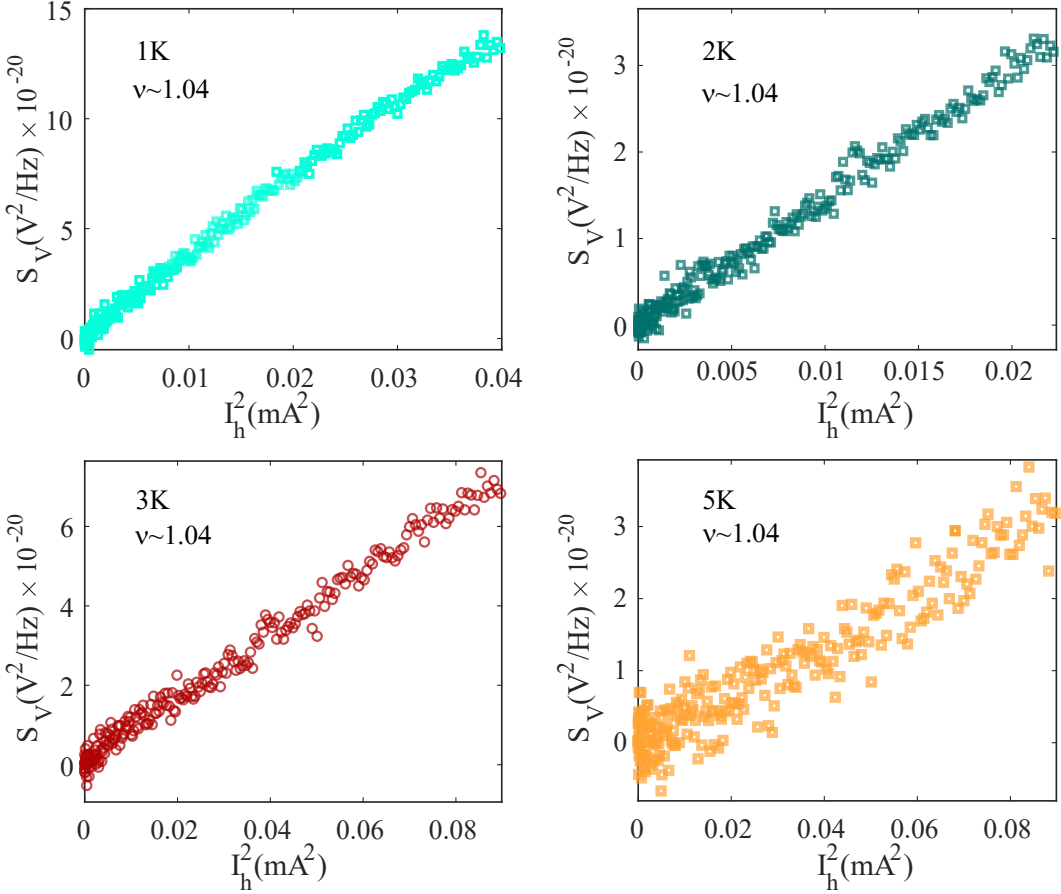

**Supplementary Figure 5:** Thermal noise ($S_V$) as function of square of applied heater current ($I^2$) for different temperatures. As can be seen $S_V$ in linear with $I^2$.

**SI-6: Gain estimation of the amplifier**

Knowing the gain of the cold amplifier-room temperature amplifier chain (described in SI. 2) is imperative to the accurate determination of $\Delta T$ and, hence thermopower. Here we use noise thermometry to measure the gain. First, we determine the resonance frequency ($f_r$) of the LC tank circuit at the input of the cold amplifier ( described in SI. 2). For this purpose we used a function generator to inject high frequency ($450-950 kHz$) signal at the source junction and measured the amplified output signal via spectrum analyzer. Supplementary SI-fig. 8(a) shows the voltage amplitude in arbitrary units of the amplifier chain as a function of measurement frequency ($f$) at $1K$ at $\nu = 1.08$. We have chosen this particular filling as the resistance ($\sim 2.5\ k\Omega$) remains invariant up to the highest measurement temperature. As can be seen from the figure, the resonance frequency occurs at $722 KHz$. Next, we measure the thermal noise by measuring $<\delta V_{out}>^2$ at the resonance frequency within a bandwidth (B.W) of $30 kHz$ as a function of varied sample temperature at the same filling at zero heater current. In SI-fig. 8(b) $<\delta V_{out}>^2 /B.W$ is plotted against temperature between $1-6K$. The plot shows the linear temperature dependence, which is expected as for any resistance R, the thermal noise is given by $S_V = 4 k_B T R = <\delta V_{out}>^2/(B.W \times A^2)$, where $K_B$ is the Boltzmann constant

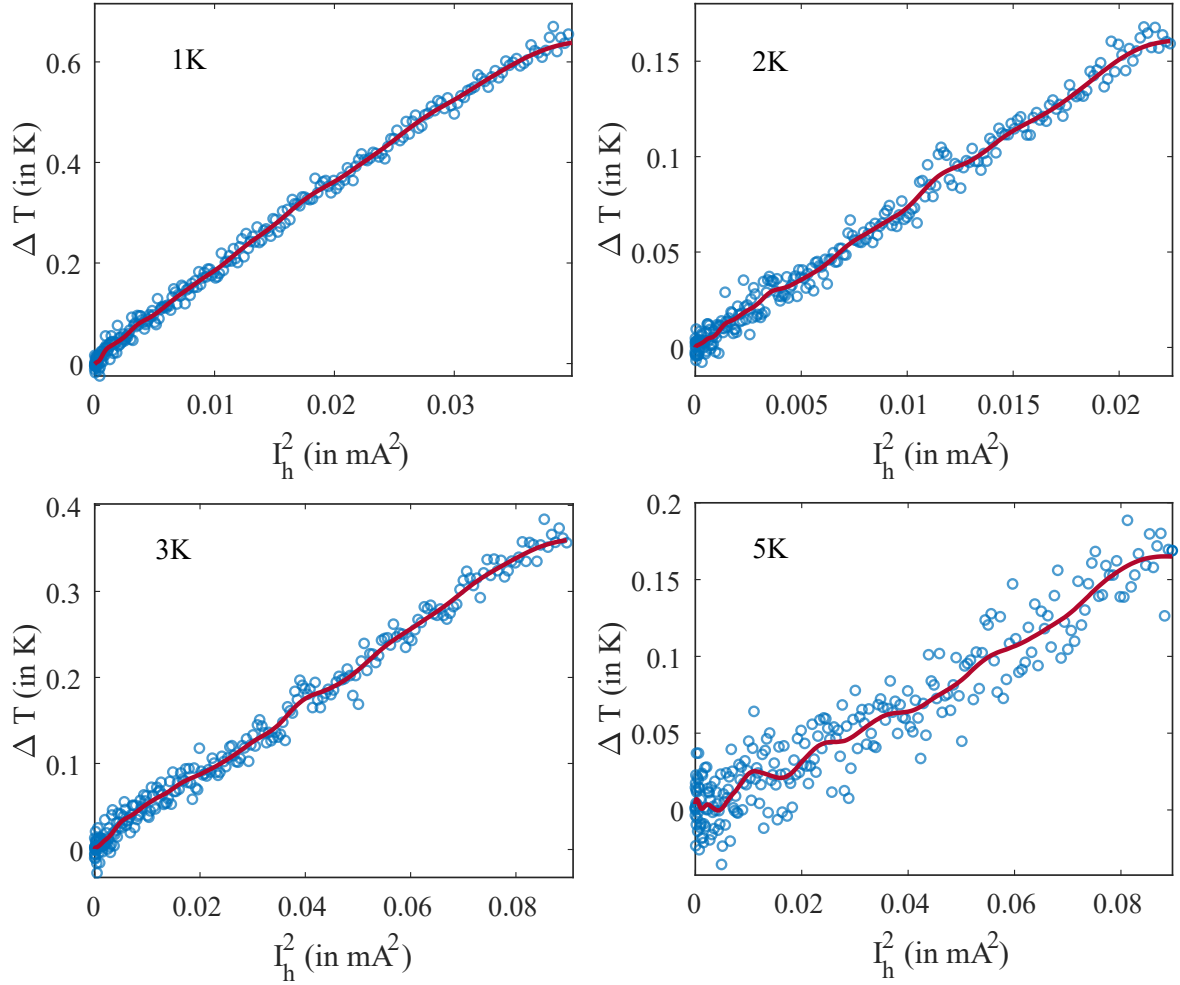

**Supplementary Figure 6:** Measured $\Delta T$ as a function of square of applied heater current ($I^2$) for different temperatures. Red line depicts the gaussian smoothening of the scattered data.

and A is the gain. The gain is determined from the slope of the plot as $A = \sqrt{\frac{slope}{4k_B R}}$ and is found to be around $\sim 1240$. Similarly at a higher temperature range (SI-fig. 8(c)) of between $12 - 24K$ the calculated gain is around $\sim 1200$ (done at a filling of $\nu = 1.03$).

### SI-7: Linear response regime of thermopower measurement

The SI-fig. 9(a) and (b) are the $V_{2\omega}$ versus back gate voltage ($V_{bg}$) for different applied heater current, respectively, at temperature 12 K and 20 K. SI-fig. 9 (c) and (d) shows the same, but $V_{2\omega}$ signal is normalized with respective heater current squared. All the normalized $V_{2\omega}/I_h^2$ values overlaps at both temperatures within the maximum applied heater current. This shows the $V_{2\omega}$ measurements have been performed in the

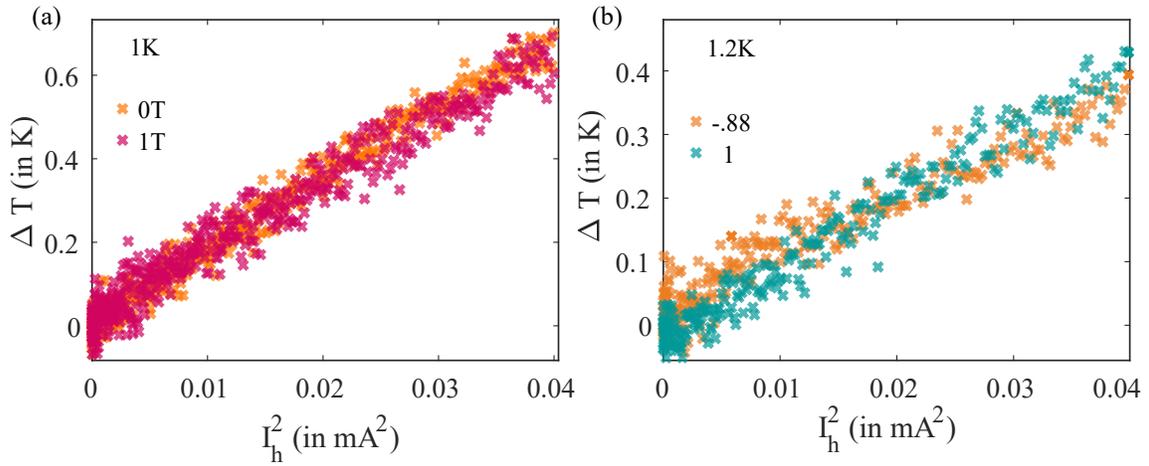

**Supplementary Figure 7:** (a)$\Delta T$ versus I$^2$ at $\nu = -3$. for zero (orange) and $B = 1T$ (red) at $T = 1K$. (b) Measured $\Delta T$ versus I$^2$ for different densities $\nu = -3.5$ (orange) and $\nu = 4$ (green) at $T = 1.2K$.

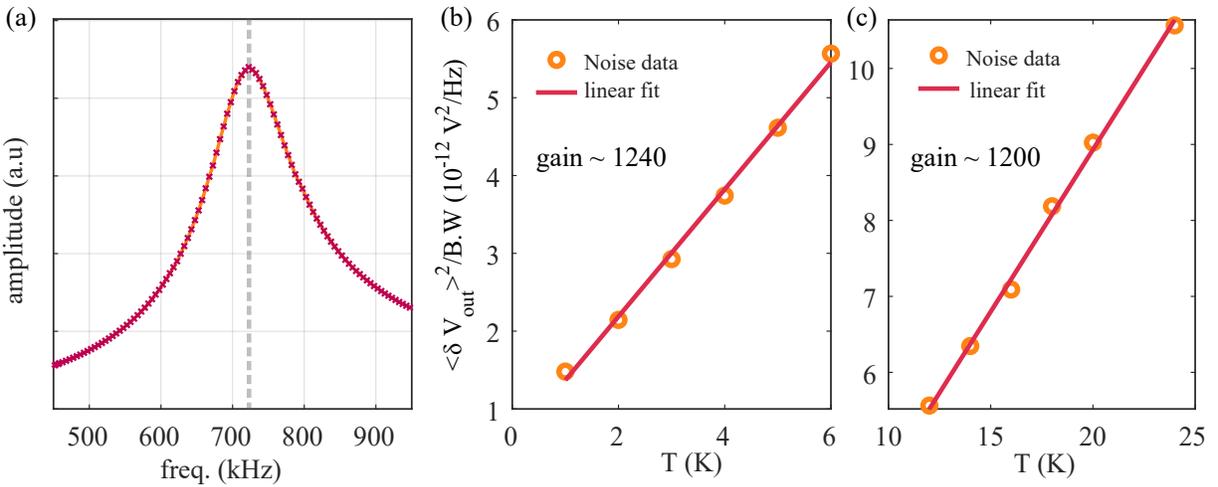

**Supplementary Figure 8:** (a) Measured output voltage (in arbitrary units) of the amplifier chain as a function of measurement frequency ($f$) for a known input signal with frequency varied from $450 - 950 kHz$, at $\nu = 0.69$. The resonance frequency ($f_r$) is $\sim 722 kHz$. (b) Measured thermal noise ($<\delta V_{out}>^2 /B.W$) as function of $T$ at $f_r$ at $\nu = 1.08$ showing linear dependence. Here B.W is the measurement bandwidth. Red line shows linear fit from which the gain ($\sim 1240$) is determined. (c) Shows a similar plot as (b) but at higher temperature range and $\nu = 1.03$

linear response regime.

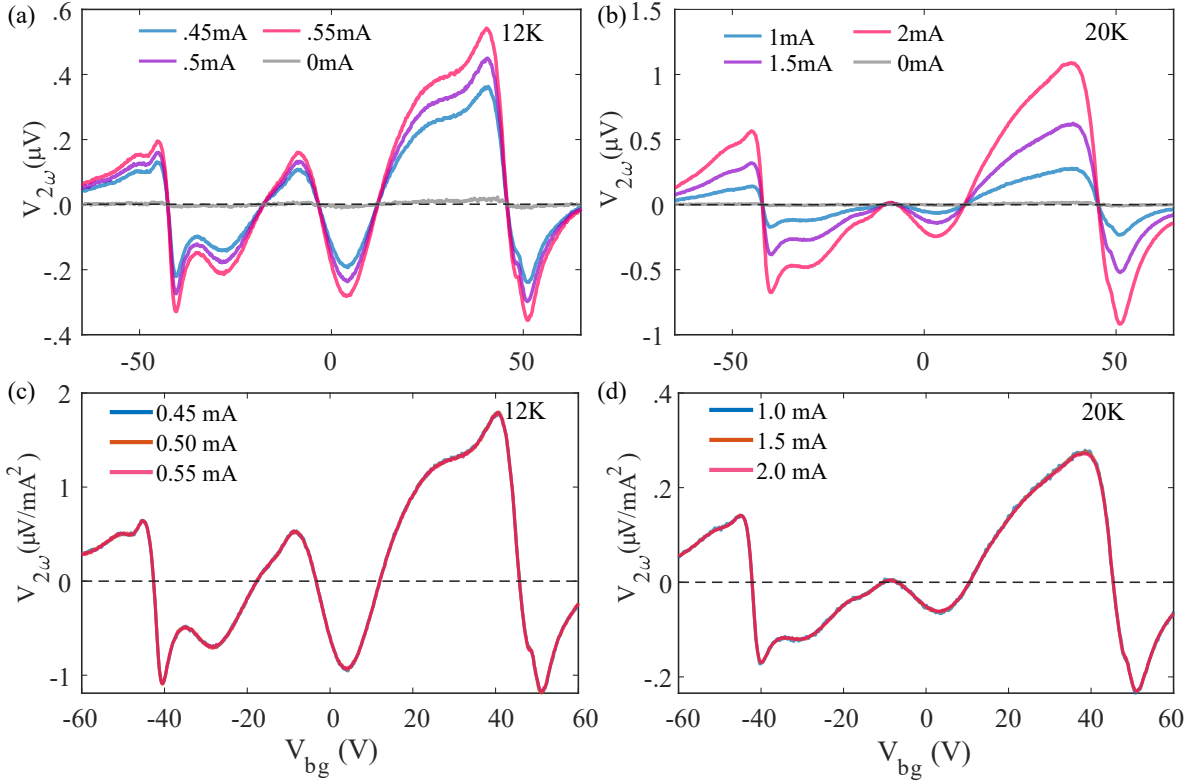

**Supplementary Figure 9:** (a) and (b) $V_{2\omega}$ versus $V_{bg}$ for different applied heater currents, respectively, at temperatures 12 K and 20 K. (c) and (d) $V_{2\omega}/I_h^2$ versus $V_{bg}$ for different applied heater currents, respectively, at mentioned temperatures.

**SI-8: Comparison of field dependent thermopower in TDBLG with MATBLG.**

SI-fig. 10 compares the evolution of thermopower near the charge neutrality in between MATBLG and TDBLG devices at $T = 1K$. SI-fig. 10 (b) and (d) show the thermopower fan in $1.05^0$(Magic angle twisted bilayer graphene) and $1.2^0$(TDBLG), respectively while (a) and (c) show the cut-lines of the same as a function of perpendicular magnetic field. The observed large enhancement of thermopower within a few hundred of mT in TDBLG is clearly absent in magic-angle twisted bilayer device. It should be noted that the MATBLG sample shows a high thermopower at CNP even at zero fields, which gradually keeps increasing with applied magnetic field at $\nu = 0$. In our previous work [2] we have addressed this thermopower peak feature to repeated reconstruction of Fermi surface at each integer filling including CNP. The gradual increase can be resulted due to modification of density of states by the application of magnetic field. For TDBLG sample in SI-fig. 10(c) we observe a further decrement in thermopower to occur beyond 2T. At these higher values of applied perpendicular magnetic field, Zeeman splitting causes a gap opening near CNP. The lack of DOS (due to this gap opening) causes a decay in thermopower with any further increase in magnetic field. At even higher fields, Landau levels start emerging (as shown in SI-fig. 10(d)) resulting in oscillations in thermopower (as displayed in the cut plot SI-fig. 10(c)) along the crossings of the Landau fans.

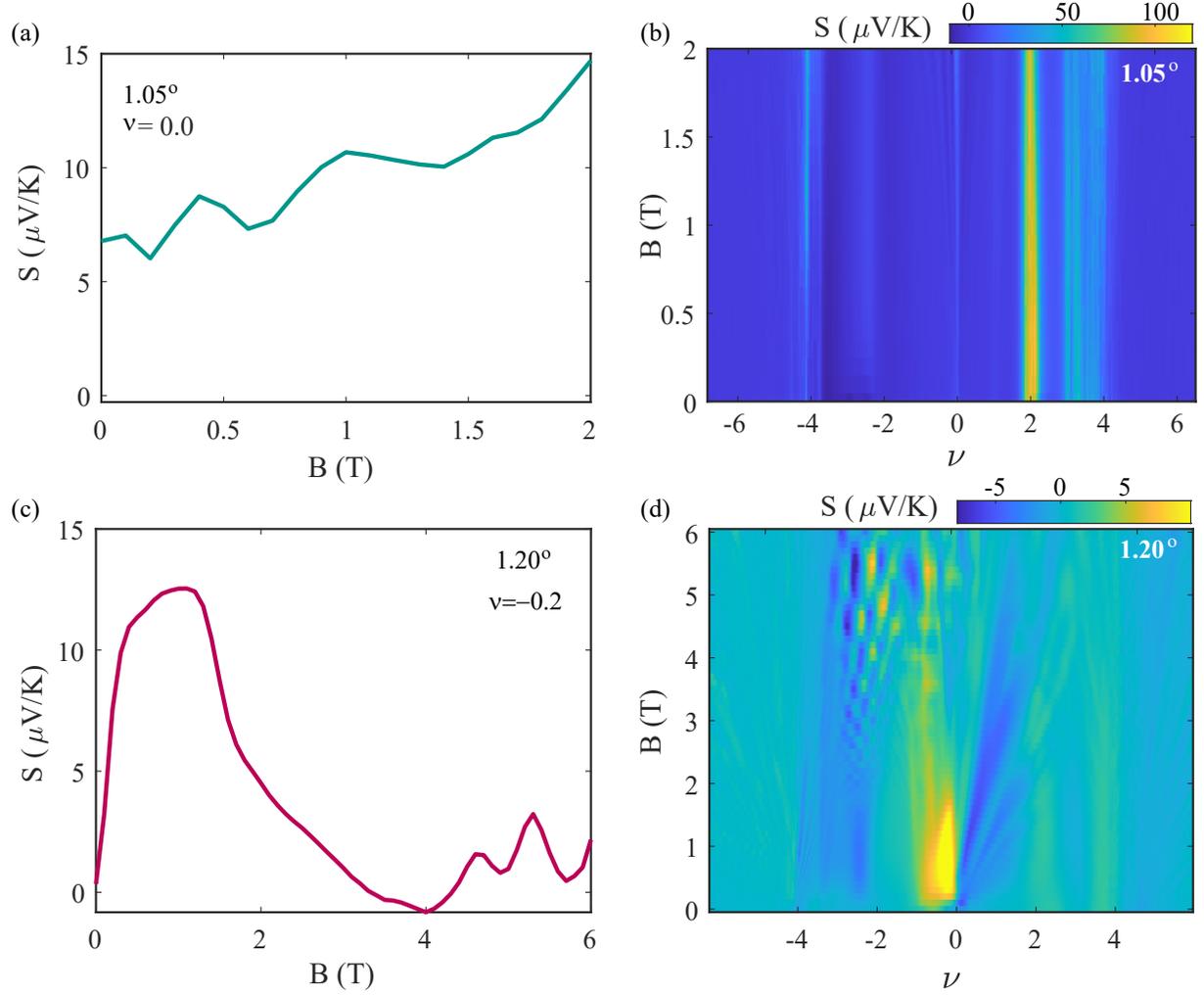

**Supplementary Figure 10: MATBLG sample:** (a) Thermopower as a function of magnetic field at $\nu = 0.0$ at $1K$. (b) 2D color plot of thermopower as a function of filling and magnetic filed at $1K$. **TDBLG sample:** (c) Thermopower as a function of magnetic field at filling close to CNP at $1K$. (d) 2D color plot of thermopower as a function of filling and magnetic filed at $1K$.

**SI-9 : Evolution of thermopower with perpendicular magnetic field at different temperature**

SI-fig. 11 shows the the magnetic field dependence of thermopower for the TDBLG at different temperatures. In the right column thermopower fans are arranged in ascending order of temperature while the left column shows corresponding thermopower as a function of magnetic field around the gate voltage (i.e density) where maximum enhancement is observed. In all temperatures form $1 - 15K$ the thermopower initially increases followed by a decrease beyond $2T$. However the peak value of the enhancement diminishes as we go beyond $5K$, and at $15K$ the enhancement reduces from $15\mu V$ to $2\mu V$. These reduction is related to the thermal broadening effect in thermopower.

10

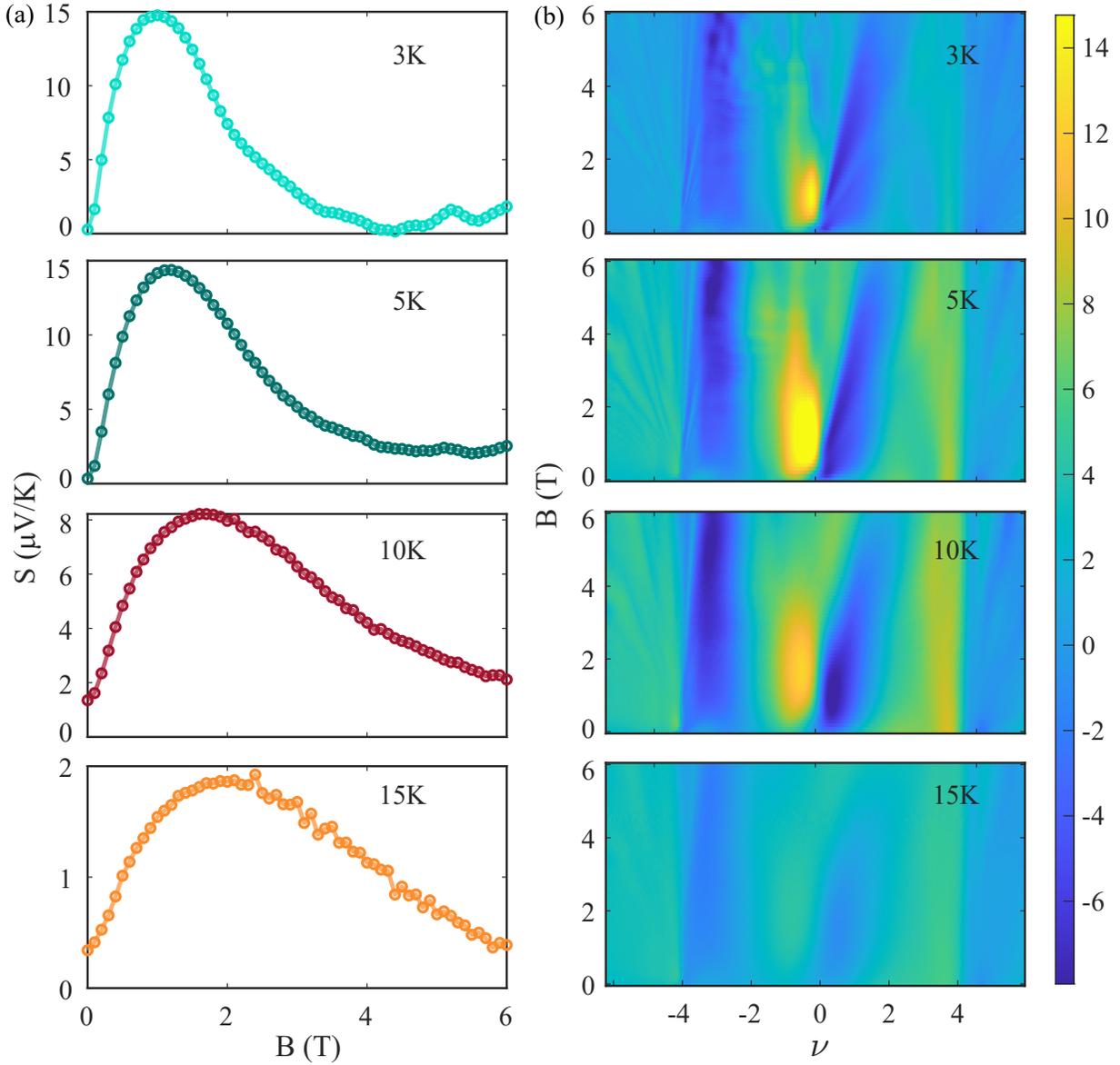

**Supplementary Figure 11:** Right column showing Seebeck fan diagram at various temperature and the left column showing corresponding vertical cuts at gate voltages where maximum enhancement is observed.

### SI-10:TDBLG Band Structure

Twisted Double Bilayer Graphene consists of two Bernal stacked (AB) bilayer graphene (BLG) sheets with a relative twist angle $\theta$ between them. Here, we work with the ABAB stacking, so that the $B$ sublattice of the top interface layer sits on top of the $A$ sublattice of the bottom interface layer.

11

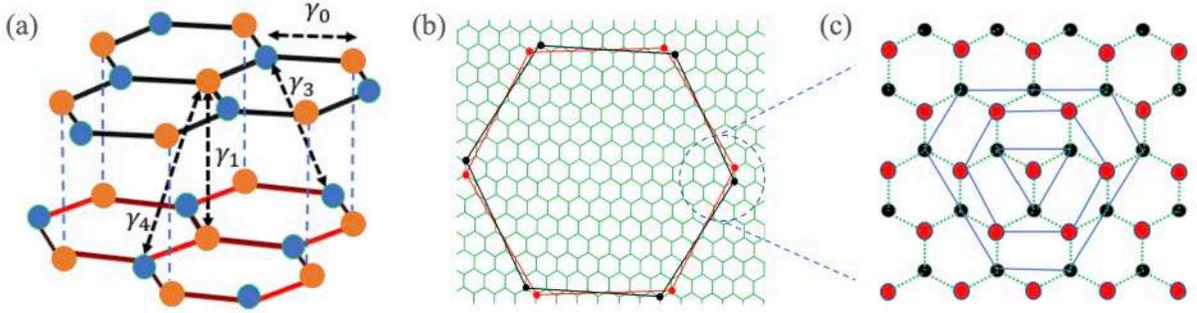

**Supplementary Figure 12:** (a) Couplings in the two layers of Bilayer graphene is schematically shown. (b) shows the tiling of the moiré Brilloiun zone (mBZ) (green hexagon) in between the two large graphene Brillouin zone(red and black hexagon) which are twisted w.r.t. each other. (c) is a zoomed in picture of the region where the continuum model of TDBLG works around the mBZ. The shell structure depicts the necessary momentum cut-off scale of the system.

Here we consider the band structure of TDBLG following Ref.s [9, 10]. The general hamiltonian for bilayer graphene can be written in the sublattice basis $(A_1, B_1, A_2, B_2)$ in the momentum space as,

$$H_{AB}(\mathbf{k}) = \begin{pmatrix} H_0(\mathbf{k}) & g^\dagger(\mathbf{k}) \\ g(\mathbf{k}) & H_0'(\mathbf{k}) \end{pmatrix} \quad (2)$$

where,

$$H_0(\mathbf{k}) = \begin{pmatrix} 0 & -\hbar v_0 k_- \\ -\hbar v_0 k_+ & \Delta' \end{pmatrix}, \; H_0'(\mathbf{k}) = \begin{pmatrix} \Delta' & -\hbar v_0 k_- \\ -\hbar v_0 k_+ & 0 \end{pmatrix} \text{ and } g(\mathbf{k}) = \begin{pmatrix} \hbar v_4 k_+ & \gamma_1 \\ \hbar v_3 k_- & \hbar v_4 k_+ \end{pmatrix}. \quad (3)$$

Here $k_\pm = \xi k_x \pm i k_y$, with $\xi = \pm 1$ for $K$ and $K'$ valley of the graphene Brillouin zone. The velocities $v_i$ are calculated from the hopping amplitudes $\gamma_i$ via, $v_i = \frac{\sqrt{3}a}{2}\gamma_i$, where $a = 0.246$ nm is the monolayer graphene lattice constant. In Fig. 12(a), we have showed the different hopping amplitudes in a bilayer graphene. The nearest neighbour tunneling amplitude along the monolayer graphene sheet, $\gamma_0$ is the strongest one out of the four. The inter-layer hopping amplitude between the $A$ sublattice in the top layer and $B$ sublattice in the bottom layer, i.e. the c-axis hopping between the dimer sites is given by $\gamma_1$ and this is the next largest scale in the problem. The inter-layer hopping between the non-dimer sites denoted by $\gamma_3$, leads to the trigonal warping of the dispersion and the weakest coupling between dimer and non-dimer sites, $\gamma_4$ breaks the particle-hole symmetry in the system. Here, $\Delta'$ is the potential difference between dimerized and non-dimerized sites and it also breaks particle-hole symmetry of the problem. For this work, we have used [9], $\hbar v_0/a = 2.1354$ eV; $\gamma_1 = 400$ meV; $\gamma_3 = 320$ meV; $\gamma_4 = 44$ meV and $\Delta' = 50$ meV.

The minimum dimension for the Hamiltonian of an ABAB stacked TDBLG is 16. This Hamiltonian

is given by

$$H_{AB-AB}(\mathbf{k}) = \begin{pmatrix} H_{AB}(\mathbf{k}, \theta/2) & U_1^\dagger & U_2^\dagger & U_3^\dagger \\ U_1 & H_{AB}(\mathbf{k}+\mathbf{q}_1, -\theta/2) & 0 & 0 \\ U_2 & 0 & H_{AB}(\mathbf{k}+\mathbf{q}_2, -\theta/2) & 0 \\ U_3 & 0 & 0 & H_{AB}(\mathbf{k}+\mathbf{q}_3, -\theta/2) \end{pmatrix}$$
(4)

The diagonal blocks are easily identified from Eq. 2, except for a relative twist of $\pm\theta/2$ from the reference hexagon which rotates the momenta as $k \to k e^{i\pm\theta}$. The off-diagonal blocks denote hopping across the twisted layers and are given by,

$$U_n = \begin{pmatrix} 0 & \tilde{U}_n \\ 0 & 0 \end{pmatrix} \quad \tilde{U}_n = \begin{pmatrix} u & u' e^{-\mathbf{i}\frac{2\pi\xi}{3}(n-1)} \\ u' e^{\mathbf{i}\frac{2\pi\xi}{3}(n-1)} & u \end{pmatrix}.$$
(5)

Here, $u$ and $u'$ are the AA/BB and AB tunneling amplitudes respectively across the twisted layers. We have used [9], $u = 79.7$ meV and $u' = 97.5$ meV for our calculations. As shown in Fig. 12 (b), the original graphene Brillouin zone is tiled by the Moire Brillouin zones (mBZ). We can take finite small number of shells which will produce reliable results for the low energy bands for these systems; e.g. the $16 \times 16$ Hamiltonian corresponds to taking only the first shell of mBZs, i.e. the red dot in the center and the three black dots surrounding it in Fig 12(c). In this work, we have taken 5 such shells which leads to a 184 dimensional matrix and gives an error of $< 1\%$ in the band dispersions at the magic angle of $1.2°$.

### SI-11: Electron Hole pockets and thermopower enhancement

Recent work by Feng et al. [11] has theoretically proposed a mechanism for the enhancement of $S$ at relatively low-magnetic fields for compensated semi-metals with electron and hole pockets. Consider a 2D system in the x-y plane with electron density $n_e$ and hole density $n_h$, where an electric field $E$ is applied along the $x$ direction and a magnetic field $B$ is applied along the $z$ direction. The longitudinal Drude electric conductivity $\sigma_{xx} = \frac{en_T\mu}{1+(\mu B)^2}$ corresponds to a drift velocity $v_d = \pm\frac{\mu E}{1+(\mu B)^2}$, while the Hall conductivity is given by $\sigma_{xy} = \frac{e\Delta n\mu^2 B}{1+(\mu B)^2}$, where $\mu$ is the common mobility of the electrons and the holes, $n_T = n_e + n_h$ is the total carrier density and $\Delta n = n_e - n_h$ is the net charge density of the carriers. In this dissipative process, the electrons and holes move in opposite directions along the applied electric field and add up their contribution to the longitudinal electric current, while they are deflected in the same direction along $y$, and hence counter each other in the Hall response. Note that this effect is reversed in heat transport, where the longitudinal heat current scales with $\Delta n$ and the transverse response scales with $n_T$. The Hall voltage along the $y$ direction leads to an electric field along this direction, given by $E_H = -\frac{\sigma_{xy}}{\sigma_{xx}}E = -\mu B \frac{\Delta n}{n_T}E$. As a result, the carriers acquire an additional spiraling component of velocity along $x$ due to the standard $\vec{E} \times \vec{B}$ drift of charged particles in crossed electric and magnetic fields. This drift velocity is in the same direction for electrons and holes and its magnitude is given by $v_{sp} = \mu E \Delta n/n_T$. Taking all the contributions into account, the electric current

$$J_e = n_T e v_d + \Delta n e v_{sp} = n_T e \mu E \left[\frac{1}{1+(\mu B)^2} + \left(\frac{\Delta n}{n_T}\right)^2\right]$$
(6)

and the heat current

$$J_Q = \frac{\pi^2}{3}\frac{k_B^2 T^2}{\epsilon_F}(\Delta n v_d + n_T v_{sp}) = \frac{\pi^2}{3}\frac{k_B^2 T^2}{\epsilon_F}\mu E \Delta n \left[\frac{1}{1+(\mu B)^2} + 1\right] \tag{7}$$

The fact that the longitudinal electric response $\sim n_T$, while the transverse response $\sim \Delta n$ opens up the possibility of having a situation where the transport is beyond the semiclassical regime ($\mu B \gg 1$), but the Hall angle $\tan\theta_H = \frac{\sigma_{xy}}{\sigma_{xx}} = \mu B \Delta n/n_T$ is still small due to the compensation from the ratio of densities. Here the electric current is still dominated by the dissipative response, $J_e \sim n_T e\mu E/(\mu B)^2$, while the heat current is dominated by the spiraling response $J_Q \sim \frac{T^2}{\epsilon_F}\mu E \Delta n$. Hence the thermopower $S_{xx} = J_Q/TJ_e$ is given by

$$S_{xx} \equiv S_1(B) = \frac{\pi^2}{3}\frac{k_B^2 T}{e\epsilon_F}\frac{\Delta n}{n_T}\mu_e^2 B^2 \tag{8}$$

This rapid quadratic rise of the thermopower is seen in our data in the main text [Fig. 2(a)]. Note that one would also expect high magnetoresistance in this regime, which is seen in the experiments. At higher magnetic fields, where $\mu B \gg n_T/\Delta n$, one enters the regime of extreme quantum transport with large Hall angles, where both the electric current and the thermal transport is dominated by the spiraling drift. In this case, the thermopower saturates as a function of the magnetic field

$$S_{xx} \equiv S_2(B) = \frac{\pi^2}{3}\frac{k_B^2 T}{e\epsilon_F}\frac{n_T}{\Delta n}. \tag{9}$$

This theoretical dependence is summarized in Fig. 2(a) inset of the main manuscript. For the region $B < B_H$ (i.e Region 1), following Eq. 8, the values for $\frac{\Delta n}{n_T}$ and $\epsilon_F$ are referred from our theoretical model and $T$ is taken $1K$. For $\mu_e$, we have chosen typical value of $10^4 cm^2 V^{-1} s^{-1}$. Note that the precise value of $\mu_e$ barely contributes to the final shape or the maximum value of the $S_{xx}$ vs. $B$ plot. Similarly, in the region $B > B_H$ (i.e Region 2), approximate values can be calculated inserting the three key input parameters $\frac{\Delta n}{n_T}$, $\epsilon_F$ and $T$. Additionally, to merge this two distinct regions of $B$ dependence in a single function we have incorporated a fermi like function $F(B - B_H)$ that gradually switches values between 1 and 0 as we cross $B = B_H$. Finally the overall Seebeck (plotted in Fig. 2(a) inset) $S(B)$ can be expressed in a single equation

$$S(B) = FS_1(B) + (1-F)S_2(B) \tag{10}$$

## SI-12: Resistance peaks and excitonic order at low temperatures:

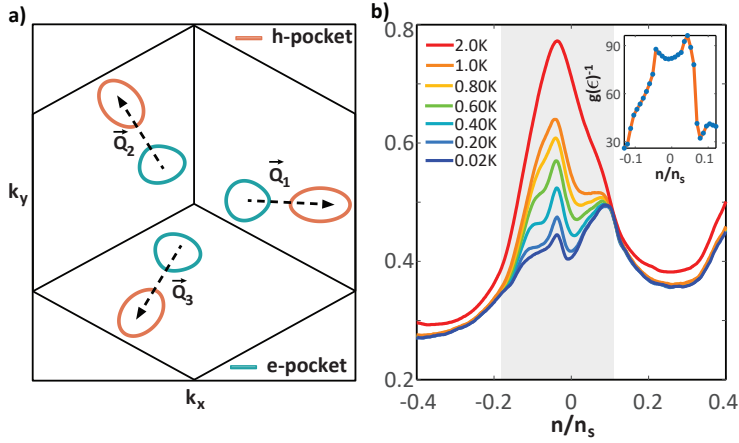

**Supplementary Figure 13:** (a) Calculated Fermi surfaces in TDBLG at CNP showing the electron (blue) and the hole (red) pockets. Note that the centres of the pockets are shifted in momentum space. (b) Low temperature resistance of TDBLG measured in absence of magnetic field as a function of carrier density $n/n_s$. Measurements at several temperatures from $20$ mK to $2\ K$ are shown. The resistance shows double peaks near CNP below $2\ K$. Electron and hole pockets coexist within the grey shaded region. The inset shows the theoretically calculated inverse density of states at the Fermi level for the excitonic metal as a function of carrier density, which shows similar two-peak features.

Fig. 13(b) shows a zoomed-in picture of resistance versus normalized density around the CNP of Fig.1(d) of main manuscript for different temperatures. Unlike any previous measurements in TDBLG, our resistance data reveals fine double peak features around CNP below $2K$. These additional features in the resistance are seen in the density range where the system shows the presence of simultaneous electron and hole pockets, as marked by the grey shaded region in the figure. Note that the total density of electron and hole states at the Fermi level of the non-interacting system is almost constant as a function of carrier density in this range and cannot explain the two peak structure of the resistance seen in the experiments. Fig. 13(a) shows the theoretically calculated Fermi surfaces in the system at CNP within the moiré Brillouin zone. We note that there are 3 electron and 3 hole pockets in the system, related by $2\pi/3$ rotations about the centre of the Brillouin zone, respecting the $C_3$ symmetry of the system. The electron and hole pockets do not overlap in momentum space; their centres are shifted by wave-vectors $\vec{Q}_{1(2)(3)}$, where $|\vec{Q}_i| \sim 0.02$ Å$^{-1}$. A detailed calculation with realistic parameters [12] shows that the Coulomb attraction between the electrons and holes in TDBLG near CNP can leads to formation of excitonic condensate. The shift between the electron and the hole pockets in momentum space may give rise to indirect excitons, which carry finite momentum (see Fig.3(d) in main text for a schematic representation); a condensate of such excitons will be equivalent to the formation of a charge density wave in the system. Note that the momenta of the excitons formed in the three electron-hole pockets add up to 0. This results in a metal with modified Fermi surfaces and excitation

spectra [12]. In the inset of Fig. 13(b), we plot the theoretically obtained inverse of the density of states at the Fermi level for this excitonic metal as a function of carrier density. This shows a two peak structure similar to the resistance curves. The correlation of the presence of the double peaks with density regions where electron and hole pockets co-exist strengthens our claim that the resistance peaks are indicative of this order.

### SI-13:The exciton condensate

The Coulomb attraction between the electron and hole pockets lead to formation of indirect exciton condensates in TDBLG near CNP. In this calculation we will replace the Coulomb potential between electrons and holes by a screened short range potential. In fact we will use an effective momentum independent potential with the energy scale $V_0 \sim 10.8$ meV. Note that there are 3 electron pockets separated from the three hole pockets by wavevectors $Q_{1(2)(3)}$. The mean field Hamiltonian describing the excitonic condensate is given by

$$\mathcal{H}(\mathbf{Q}_i) = \begin{bmatrix} \frac{1}{3}\epsilon_\mathbf{k}^c - \mu & \Delta \\ \Delta & \frac{1}{3}\epsilon_{\mathbf{k}+\mathbf{Q}_i}^v - \mu \end{bmatrix} \quad (11)$$

where, the $\epsilon_\mathbf{k}^{c(v)}$ represents the non-interacting conduction (valence) band dispersion and the chemical potential is denoted as $\mu$. Note the Hamiltonian is a $6 \times 6$ matrix which can be easily block diagonalized into three $2 \times 2$ matrices, one for each $\mathbf{Q}$ [12]. The order parameter $\Delta$ is same for all the pockets and is determined self-consistently. We can then write the modified quasi-particle dispersion relation in presence of the excitonic condensate,

$$E_{\mathbf{Q}_i}^\pm(k) = \frac{\epsilon_\mathbf{k}^c + \epsilon_{\mathbf{k}+\mathbf{Q}_i}^v}{6} - \mu \pm \sqrt{\frac{(\epsilon_\mathbf{k}^c - \epsilon_{\mathbf{k}+\mathbf{Q}_i}^v)^2}{36} + \Delta^2} \quad (12)$$

The above energy spectrum generates a finite Fermi surface near CNP, which leads to metallic transport in presence of the condensate. The modified density of states at the Fermi level can then be calculated from

$$\rho(\mu) = \sum_{\mathbf{k},i} \delta(E_{\mathbf{Q}_i}^\pm(\mathbf{k}) - \mu) \quad (13)$$

We use the inverse DOS ($\rho^{-1}(\mu)$) as a proxy for the resistivity of the material and this inverse DOS is plotted as a function of carrier density in the SI-12 [SI-fig 13 inset] to indicate the rough behaviour of the resistance in the system near CNP.

# SI-14: The Comsol simulations of linear temperature profile

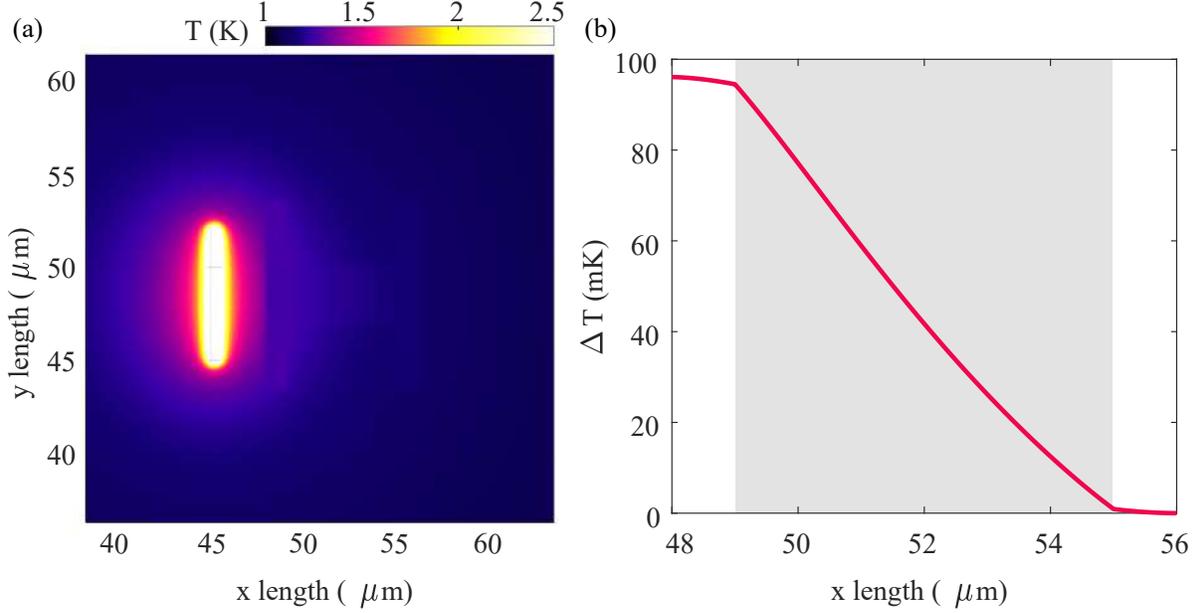

**Supplementary Figure 14:** (a) 2D color plot of the simulated temperature profile across the XY-plane on $SiO_2$ surface. (b) Temperature difference across graphene plane (X-direction) showing linear temperature profile along the graphene channel (grey shaded region).

In section SI-4, we have assumed a linear fall off of the temperature profile across the double bi-layer graphene channel, which allows us to express the excess thermal noise, $S_V = 2k_B \Delta T R$. We have solved three dimensional Fourier heat diffusion equations for a multi-layer stack using finite element calculations in Comsol to test the validity of such assumptions. We find the simulated temperature profile to be just about linear and entirely in line with our assumptions.

For the purpose of simulation, the multi-layer stack has been modelled after the actual device geometry and dimensions. The model consists of top $hBN$ ( 30 nm), twisted double bilayer graphene ( 10 nm), bottom $hBN$ ( 30 nm), $SiO_2$ ( 300 nm), $Si$ ( 100 µm). Note that we have considered a 10 nm thick graphite sheet in our simulation model instead of $\sim$ 1nm thick twisted bilayer graphene due to the computational limitation of fineness of grid size. However, we have adjusted the thermal conductivity of the graphene accordingly to counter the effects of the extra thickness. The length (L) and width (W) of $hBN$/Graphite flake/$hBN$ are 6 µm and 3 µm, respectively. The $Si/SiO_2$ substrate is 100 µm × 100µm. The $SiO_2$ film being 300nm thick. The gold probes and the heater dimensions are 10 µm(L) ×1 µm(W) × 0.07 µm(d) and 7 µm(L) × 0.3 µm(W) × 0.07 µm(d), respectively, and the heater is placed at 3 µm away from the left gold contact i.e the source contact.

The key input parameters for the simulation are, thermal conductivity ($\kappa$) of each material, the ther-

17

mal boundary conductance ($G$) between the interfaces, the heater current and the boundary conditions for the temperatures are the key input parameters. We have referred to preexisting literature values for the thermal conductivity of each materials at $\simeq 1K$. At low temperature, the in plane thermal conductivity $\kappa_{xy}$ for $hBN$ is thickness-independent and is found to be $3Wm^{-1}K^{-1}$ at 1K [13]. In graphene, considering both the electronic and lattice contributions, we have taken total thermal conductivity $\simeq 2Wm^{-1}K^{-1}$ at $1K$ [5, 14]. But, in order to get the realistic results with a $10nm$ thick graphite flake instead of $\simeq 1nm$ twisted double bilayer graphene, we have used the reduced value to $\simeq 0.2Wm^{-1}K^{-1}$. For $SiO_2$ and $Si$, we refer to previous low temperature measurements [15, 16] and we have used $\kappa_{SiO_2} \simeq 0.01Wm^{-1}K^{-1}$, and $\kappa_{Si} \simeq 0.1Wm^{-1}K^{-1}$ at $1K$. The thermal conductivity for the gold contacts [17] of $\simeq 20Wm^{-1}K^{-1}$ at $1K$ is used in our simulation. The interfaces are known to limit heat transport in vdW heterostructures. For graphene/$hBN$, we refer to calculations based on non-equilibrium Green's function [18]; the boundary thermal conductance for $hBN$/graphene $\simeq 1MWm^{-2}K^{-1}$ at $5K$. The same boundary thermal conductance is assumed for the $hBN/SiO_2$ and $SiO_2/Si$ interfaces, and kept it same for our desired temperature of $1K$, since the introduction of boundary conductance do not significantly affect the temperature profile. For the simulation, we use the powers (P) dissipated in the heater $1\mu W$ at $1K$.

In SI-fig 14 we present the simulated temperature profile at $1K$. A 2d color plot of the temperature profile in the $SiO_2/Si$ is shown in SI-fig 14(a). In the panel SI-fig 14(b) the temperature cut line along the graphene channel is shown. Along the length of the channel (marked in grey shade), a fairly linear decay-profile of temperature is evident, corroborating our conjecture of the linear temperature profile in deriving the temperature difference using $S_V = 2k_B \Delta T R_s$ as mentioned in SI-4.


\end{document}